\documentclass[journal=tches]{iacrtrans}

\usepackage{amsmath}
\usepackage{amssymb}
\usepackage{graphicx}
\usepackage{caption}
\usepackage{subcaption}
\usepackage{xcolor}
\usepackage{booktabs}
\usepackage[hidelinks]{hyperref}

\usepackage{stmaryrd}
\usepackage{multirow}

\usepackage{wasysym}
\usepackage{threeparttable}

\newcommand{\schemename}{\textsc{Dash}}
\newcommand{\slalom}{\textsc{Slalom}}
\newcommand{\gazelle}{\textsc{Gazelle}}
\newcommand{\gnnp}{\textsc{gnnp}}
\newcommand{\delphi}{\textsc{Delphi}}
\newcommand{\secureml}{\textsc{SecureML}}
\newcommand{\minionn}{\textsc{MiniONN}}
\newcommand{\piranha}{\textsc{Piranha}}
\newcommand{\cryptgpu}{\textsc{CryptGPU}}
\newcommand{\cryptflow}{\textsc{CrypTFlow}}
\newcommand{\mptwoml}{\textsc{MP2ML}}
\newcommand{\deepsecure}{\textsc{DeepSecure}}
\newcommand{\simc}{\textsc{SIMC}}
\newcommand{\muse}{\textsc{Muse}}
\newcommand{\goten}{\textsc{Goten}}
\newcommand{\etal}{et~al.\ }

\usepackage{fixme}
\usepackage{fontawesome}
\usepackage{enumitem}

\usepackage{listings}
\usepackage{array}

\author{Jonas Sander\inst{1} \and Sebastian Berndt\inst{2} \and Ida Bruhns\inst{1} \and Thomas Eisenbarth\inst{1}}
\institute{
	University of Luebeck, Luebeck, Germany, \email{{j.sander,ida.bruhns,thomas.eisenbarth}@uni-luebeck.de}
	\and
	Technische Hochschule Luebeck, Luebeck, Germany, \email{
sebastian.berndt@th-luebeck.de}
}

\title[Dash: Accelerating Private Inference with Arithmetic Garbled Circuits]{Dash: Accelerating Distributed Private Convolutional Neural Network Inference with Arithmetic Garbled Circuits}

\begin{document}

\maketitle


\begin{abstract}
The adoption of machine learning solutions is rapidly increasing across all parts of society.
As the models grow larger, both training and inference of machine learning models is increasingly outsourced, e.g. to cloud service providers. This means that potentially sensitive data is processed on untrusted platforms, which bears inherent data security and privacy risks. 
In this work, we investigate how to protect distributed machine learning systems, focusing on deep convolutional neural networks. The most common and best-performing mixed MPC approaches are based on HE, secret sharing, and garbled circuits. They commonly suffer from large performance overheads, big accuracy losses, and  communication overheads that grow linearly in the depth of the neural network. To improve on these problems, we present \schemename{}, a fast and distributed private convolutional neural network inference scheme secure against malicious attackers.
Building on arithmetic garbling gadgets \cite{DBLP:conf/ccs/BallMR16} and  \emph{fancy-garbling} \cite{DBLP:journals/iacr/BallCMRS19}, \schemename{} is based purely on arithmetic garbled circuits. 
We introduce \emph{LabelTensors} that allow us to leverage the massive parallelity of modern GPUs. 
Combined with state-of-the-art garbling optimizations, \schemename{} outperforms previous garbling approaches up to a factor of about 100.
Furthermore, we introduce an efficient \emph{scaling} operation over the residues of the Chinese remainder theorem representation to arithmetic garbled circuits, which allows us to garble larger networks and achieve much higher accuracy than previous approaches.
Finally, \schemename{} requires only a single communication round per inference step, regardless of the depth of the neural network, and a very small constant online communication volume. 
\end{abstract}

\section{Introduction}
The recent progress of machine learning (ML) has resulted in broad adoption across almost all industries and government institutions.
Consequently, ML is also being applied to security and privacy-sensitive domains like healthcare, law enforcement, finance, public administration, logistics, and many more. As ML models usually become very large and computation intensive, machine learning as a service (MLaaS) is a growing market in which data owners send their data to an inference device (or server) which produces the output. However, the owners (often called clients) of input and output may not want to or are not allowed to share their data due to privacy concerns or due to regulations such as the General Data Protection Regulation (GDPR) of the European Union. This can be solved by providing private inference protocols. 

Many different cryptographic approaches have been used to protect the data in distributed ML applications, including secure multiparty computation (MPC)~\cite{DBLP:conf/iclr/TramerB19}, homomorphic encryption (HE)~\cite{DBLP:journals/corr/abs-1811-09953},  garbled circuit (GC) based techniques~\cite{DBLP:conf/dac/RouhaniRK18}, or combinations of the former~\cite{DBLP:conf/sp/MohasselZ17,DBLP:conf/uss/JuvekarVC18,DBLP:conf/uss/LehmkuhlMSP21,DBLP:conf/uss/Chandran0OS22}. While HE and MPC already provide quite efficient solutions for linear operations, the non-linear components of artificial neural networks (ANNs) lead to very large overheads in these schemes. Newer approaches thus split up the operations of ANNs and compute the linear components via HE or MPC and the non-linearities via GCs \cite{DBLP:conf/asiacrypt/ChangL01, DBLP:conf/mmsec/BarniOP06, DBLP:journals/ejisec/OrlandiPB07, DBLP:journals/tifs/BarniFLS011, DBLP:conf/sp/MohasselZ17, DBLP:conf/ccs/LiuJLA17, DBLP:conf/ccs/RiaziWTS0K18, DBLP:conf/ccs/MohasselR18, DBLP:conf/uss/MishraLSZP20}.
Unfortunately, switching the cryptographic techniques seems to always require communication between the data owner(s) and the inference device or between multiple inference devices which jointly perform the inference. This results in a linear number of communication rounds in the number of transitions from linear to non-linear layers (and vice versa) in the model architecture. Consequently, the computation of a linear layer that follows a non-linear layer must always wait for the completion of the required communication, slowing down the computation significantly.
Even worse, in many approaches, the communication volume massively grows with the depth of the network~\cite{DBLP:conf/ccs/LiuJLA17,DBLP:journals/corr/abs-1811-09953}, turning communication volume into a massive bottleneck compared to the computational cost of the schemes.
Excessive volumes of communication can be avoided by staying within a single technique, as long as all layer types can be computed efficiently within the chosen technique.
Compared to other MPC techniques, GCs require the least amount of online communication.
Several works have proposed and analyzed the usage of binary GCs for protecting machine learning~\cite{DBLP:conf/icisc/SadeghiS08, DBLP:conf/dac/RouhaniRK18, DBLP:conf/uss/RiaziS0LLK19}. 
While keeping online communication rounds low, binary GCs are not well-suited for the arithmetic of ANNs, resulting in huge and slow circuits. 
In their fairly recent seminal work, Ball et~al. generalized the concept of GCs to  \emph{arithmetic GCs}~\cite{DBLP:conf/ccs/BallMR16}, which can be tailored to accommodate ANN-specific arithmetic \cite{DBLP:journals/iacr/BallCMRS19}, significantly outperforming previous binary GC approaches. \schemename{} builds on the cryptographic building blocks of these works, extends them with efficient power-of-two scaling in the Chinese remainder theorem representation, demonstrates their applicability in the offline-online scenario and supports GPU acceleration.

Non-private ML settings massively benefited from the use of accelerators like GPUs. Hence, using acceleration-friendly techniques for distributed private inference (DPI) is a logical step. While binary GCs were accelerated on GPUs already a decade ago \cite{DBLP:conf/acsac/HustedMSG13}, to the best of our knowledge, \schemename{} is the first framework demonstrating GPU acceleration of arithmetic GCs. In modern works on accelerating cryptographic inference using GPUs such as GForce \cite{DBLP:conf/uss/NgC21}, GPU-accelerated GCs are not common. In fact, the authors \cite{DBLP:conf/uss/NgC21} formulate efficient GPU acceleration of GCs as an open challenge. \schemename{} accelerates highly optimized arithmetic GCs for CNN inference including the new garbled scaling gadget on the GPU and outperforms the previous state-of-the-art in outsourced inference.

An alternative to purely cryptographically protected distributed inference is to rely on TEEs such as Intel SGX or AMD SEV for the protection of data and/or model~\cite{DBLP:journals/corr/abs-2208-10134,DBLP:conf/aaai/NgCWW021}. \slalom{} used a TEE combined with a simple stream cipher to protect input data during inference while leveraging a co-located GPU to accelerate linear layers~\cite{DBLP:conf/iclr/TramerB19}. Combining lightweight cryptography with efficient TEE solutions results in small runtimes for \slalom{}. However, \slalom{} needs to communicate the entire layer output between TEE and GPU after each switch from linear to non-linear layer (and the other way around), making the GPU's co-location to the TEE mandatory. \schemename{} can also use a TEE but only needs to communicate the model in- and outputs with the inference device (e.g., GPU accelerated) at online time, making the computation of the hidden layers non-interactive.

\subsection*{Our contribution}
In this work, we propose a novel arithmetic garbling framework based on the ANN-optimized arithmetic GCs from Ball et~al. \cite{DBLP:journals/iacr/BallCMRS19}.
\schemename{} uses heavily optimized GCs to offer strong security guarantees for confidentiality of the inputs and outputs and the integrity of the computation even in the presence of malicious parties.
To achieve security against such malicious parties, we can either rely on purely cryptographic techniques such as cut-and-choose and zero-knowledge proofs or use TEEs.
Furthermore, the whole ANN can be garbled securely while the performance-critical GC evaluation (the protected ANN inference) can be performed by an untrusted device. 
We introduce the notion of \emph{LabelTensors} that allow massive acceleration of this inference on GPUs, resulting in significantly improved performance during inference compared to previous solutions. 
\schemename{} thus combines the advantages presented by GCs (very low communication complexity) with the advantages presented by GPUs (massive  parallelism).  
Furthermore,  we introduce an efficient \emph{scaling} operation to arithmetic GCs, which allows us to further push the 
limit of the size of ANNs that can be efficiently garbled and thus the accuracy of the inference.  
Compared to secret-sharing-based MPC frameworks like \piranha{}~\cite{DBLP:conf/uss/WatsonWP22}, where each party contributes a share and the memory requirement increases linearly, our solution has a constant memory requirement independent of the number of input providers.
In summary, our contributions are:
\begin{itemize}[noitemsep,topsep=0pt]

    \item Introducing \emph{LabelTensors} which leverage the intrinsic parallel nature of ANNs and allows us to evaluate them very efficiently on highly parallel GPUs. As a result, \schemename{} outperforms inherently serial state-of-the-art solutions.
    \item Enabling modern quantization schemes by developing a \emph{scaling} operation to achieve much higher accuracy on larger networks than previous GC frameworks as well as increasing the ANN size a GC can  handle effectively. 
    \item Regardless of the number of layer-transitions from linear to non-linear and vice versa in the model architecture, \schemename{} requires only a constant number of communication rounds and a constant communication volume.
        \item Combining GCs with TEE-enabled modern hardware to guarantee \emph{input privacy}, \emph{integrity of the computation}, and \emph{output privacy}.
        \item \schemename{} can also be extended to guarantee security in the presence of an actively \emph{malicious} attacker controlling multiple parties.
    \item \schemename{} has a constant computation-, memory-, and com\-mu\-ni\-ca\-tion\--over\-head independent of the number of participating input owners.
    \item \schemename{} is open-source and source code will be released upon acceptance.\footnote{\url{https://github.com/UzL-ITS/dash}}

\end{itemize}

\section{Related Work}
In 2017, Mohassel and Zhang \cite{DBLP:conf/sp/MohasselZ17} introduced \secureml{}, a scheme for private inference and training in a setting with two untrusted but \emph{non-colluding} servers, which then train the desired model using 2-party MPC techniques. The activation functions sigmoid and softmax are approximated due to their high complexity in an MPC setting, and GCs are used to further speed up the computation of these functions.
\deepsecure{}~\cite{DBLP:conf/dac/RouhaniRK18} preprocesses the networks and encodes them into a \emph{binary} circuit for garbling. During the evaluation, optimized oblivious transfers are used for the evaluation of non-linear operations. 
Liu \etal introduced \minionn{} for secure inference on ANNs in the one-server setting~\cite{DBLP:conf/ccs/LiuJLA17}. It also approximates activation functions and combines secret sharing with GCs. The evaluation of all these schemes shows that ANNs with seven or more activation layers generate so much overhead that the scheme is not usable any more. 

\gazelle{} builds on \minionn{} and uses HE in the linear layers and GCs in the activation layers \cite{DBLP:conf/uss/JuvekarVC18}. Every part of \gazelle{} was optimized for performance, including evaluating the linear layers and transforming the data between the linear and non-linear layers. 
\gazelle{} supports inference only and advanced the state-of-the-art runtime at the time by several orders of magnitude, mainly by optimizing the  communication.  

All these schemes use the CPU only. Since ML has benefited from GPUs, it is a natural direction to explore when trying to improve the performance of private ML schemes. \delphi{} \cite{DBLP:conf/uss/MishraLSZP20} transferred the techniques used by \gazelle{} to the GPU~\cite{DBLP:conf/uss/MishraLSZP20}. It accelerates the linear layers by moving the HE computations via additive secret sharing into the offline phase. In addition, a performance-accuracy tradeoff for $\operatorname{ReLUs}$ is calculated and hyperparameter optimization is used for the resulting architecture.
This architecture remodeling of the ANN is timely and requires extensive retraining of the model.

Continuing with the GPU approach, Tram\`er and Boneh introduced the \slalom{} scheme in 2019~\cite{DBLP:conf/iclr/TramerB19}, providing verified and private inference on ANNs in a one-server setting. The server must feature a CPU with a TEE and a co-located GPU.
Several attempts to secure the one-server by leveraging a TEE have been introduced \cite{DBLP:conf/uss/OhrimenkoSFMNVC16, DBLP:journals/corr/abs-1810-00602, DBLP:journals/corr/abs-1803-05961, DBLP:journals/corr/abs-1808-00590, DBLP:conf/mobicom/LeeLPLLLXXZS19, DBLP:journals/ieeesp/ZhangHCKHJJMS20}, but \slalom{} is the first protocol to privately outsource the expensive linear operations of ANN inference from the TEE to a GPU via masking. The communication overhead increases linearly with the depth of the ANN. Like \schemename{}, \slalom{} can also protect against a malicious attacker. 

Faster CryptoNets~\cite{DBLP:journals/corr/abs-1811-09953} is a scheme for encrypted inference in the one-server setting which follows the two previously proposed CryptoNets \cite{DBLP:conf/icml/Gilad-BachrachD16, DBLP:journals/corr/XieBFGLN14} approaches and uses HE to perform ANN inference over encrypted inputs. The activations are approximated by quantized polynomials, which works best in the small interval $[-1,1]$. Outside this interval, more significant errors occur. The practicable multiplicative depth of the HE scheme limits the ANN to three layers, after which the authors delegate some computations back to the clients, which contradicts the MLaaS setting and induces a large communication overhead of several hundred GBs to TBs.

Recent approaches, such as \piranha{} in 2022, focus on the usability and performance of the suggested solution~\cite{DBLP:conf/uss/WatsonWP22}. The \piranha{} platform allows developers of secret-sharing-based MPC schemes to use GPUs efficiently without knowledge of GPU programming. The models do not need to be retrained. Besides that \piranha{} does not add any features to the existing schemes. \schemename{} also focuses on usability: Model and input owners can use the framework without knowledge of GCs or TEEs. They do not need to modify their ML Model since \schemename{} supports loading models in ONNX format. 

\section{Preliminaries}\label{sec:prelim}
We introduce the basics of ANNs, GCs, and TEEs to facilitate the explanation of \schemename{}. 

\subsection{Neural Networks}
For classification tasks, ANNs are used to map an input to an output-class, e.g., a picture of a handwritten digit to the output $0$ - $9$.  Usually, this is done by a variety of layers that are consecutively applied to the input. 
For our purposes, we will consider \emph{linear layers} such as dense layers and 2D convolutions, and \emph{non-linear layers} like $\operatorname{ReLU}$ and $\operatorname{sign}$ activations.

\subsection{Garbled Circuits}
GCs were introduced by Yao \cite{DBLP:journals/iacr/Rabin05} and allow two-party MPC computations of binary circuits against a semi-honest or malicious attacker \cite{DBLP:conf/eurocrypt/LindellP07}. To garble a gate with two input wires $x$ and $y$ and an output wire $z$, the garbler generates two random labels $l^0, l^1 \in_r \mathbb{Z}_2^\kappa$ of lengths $\kappa$ representing the 0- and 1-bit semantic for all in- and outputs. For a given gate functionality $f\colon \mathbb{Z}_2^2 \rightarrow \mathbb{Z}_2$, the garbler produces the garbled gate as a table of 4 ciphertexts $\operatorname{EN}_{l_{x}^a,l_{y}^b}(l_{z}^{f(a,b)})$ for all $a,b\in \mathbb{Z}_2$.
Here, $l_{u}^a$ corresponds to the label of wire $u$ with semantic $a$ and 
$\operatorname{EN}_k$ is a suitable encryption function with key $k$.

To garble an acyclic circuit of several gates, the above procedure is applied successively from the input to the output gates. Since the evaluator only knows the wire labels for a single input combination, he learns only the output label corresponding to this input. To prevent the evaluator from distinguishing between labels with 0- and 1-bit semantics, GCs can only be used for a single run. Besides, the ciphertexts must be shuffled per gate, as a canonical ordering  exposes the key-bit-mapping. In the classical setting, the evaluator uses an oblivious transfer to obtain the labels of his private inputs.

We follow the conventional notation and note the garbling algorithm, which generates the garbled circuit $\mathsf{gC}$ from a given circuit $\mathsf{C}$, as $\mathsf{Gb(C)} = \mathsf{gC, e, d}$. The encoding information $\mathsf{e}$ are used by the encoding algorithm to garble the clear input and obtain the garbled input $\mathsf{En(In, e) = gIn}$. We note the evaluation algorithm, which evaluates a garbled input on a GC and outputs the garbled output as $\mathsf{Ev(gC, gIn) = gOut}$. To decode the garbled output, the decoding algorithm is used with the decoding information $\mathsf{De(gOut, d) = Out}$. As \schemename{} deals with neural networks, we also write $\mathsf{NN}$ and $\mathsf{gNN}$ instead of $\mathsf{C}$ and $\mathsf{gC}$.

\subsection{Garbled Circuit Optimizations}
Optimizations for GCs focus on their size, the computational complexity, and the hardness assumptions needed to achieve appropriate security guarantees. Below, we concentrate on optimizations which generalize to the arithmetical domain and are supported by \schemename{}.

\textit{Point-and-permute} was introduced by Beaver \etal \cite{DBLP:conf/stoc/BeaverMR90} and reduces the number of necessary decryption operations to one ciphertext per gate. The idea is to append a pointer pair $(p, \overline{p})$ with $p \in_r \mathbb{Z}_2$ randomly chosen to each pair of input labels $(l^0\mathbin\Vert p, l^1\mathbin\Vert \overline{p})$ to sort the ciphertexts based on these pointer (called color bit). This ordering allows the evaluator to determine the correct output using only one decryption. 
\textit{Free-XOR} was introduced by Kolesnikov and Schneider~\cite{DBLP:conf/icalp/KolesnikovS08} and enables the evaluation of XOR gates without performing cryptographic operations or transmitting ciphertexts. The garbler chooses the input labels $l^0$ with 0-bit semantics randomly, and the labels with 1-bit semantics as $l^1=l^0 \oplus R$, with $R \in_r \mathbb{Z}_2^k$ being a circuit-wide constant. 
With output label $l_z^0 = l_x^0\oplus l_y^0$, the output of an XOR gate is simply evaluated as the XOR of the two input labels. 
\textit{Half Gates} garble AND gates with only two ciphertexts per gate~\cite{DBLP:conf/eurocrypt/ZahurRE15}. We give a detailed description of the arithmetical generalization used in \schemename{} below.

The optimizations described so far mainly focus on the communication complexity of GC-based protocols.
Another line of work also introduces computational improvements regarding the gate-level encryptions \cite{DBLP:conf/sigecom/NaorPS99, DBLP:conf/scn/LindellPS08, DBLP:conf/uss/HuangEKM11, DBLP:conf/uss/KreuterSS12}.  The state-of-the-art was presented by Bellare \etal \cite{DBLP:conf/sp/BellareHKR13}. They propose to encrypt wire labels using a cryptographic permutation instantiated by fixed-key \textsc{AES}. By instantiating \textsc{AES} with a fixed and public key, the scheme only needs to perform a single key-derivation for a whole circuit. While avoiding key derivations drastically improves the computation time, it comes at the cost of introducing non-standard assumptions about \textsc{AES} to enable a security-proof in the random-permutation model (for a more details, see \cite{DBLP:conf/ccs/GueronLNP15, DBLP:conf/sp/GuoKW020}).

\subsection{Arithmetic Garbled Circuits}
\label{sec:arithmetic_garbled_circuits}
Implementing arithmetic operations via conventional binary GCs is expensive, especially compared to other MPC approaches, such as secret-sharing-based MPC. Ball, Malkin, and Rosulek~\cite{DBLP:conf/ccs/BallMR16} introduced garbling gadgets for efficient garbling of arithmetic circuits over large finite fields.
We will refer to their approach as BMR scheme. Considering our use case, their gadgets allow free addition, free multiplication with a constant, and efficient projection gates for arbitrary unary functions (w.r.t.~communication complexity).

Starting from Free-XOR, Ball \etal \cite{DBLP:conf/ccs/BallMR16} consider labels as vectors of components
from $\mathbb{Z}_p$. The encoding of a value $a \in \mathbb{Z}_p$ is given through $l^a = l^0 + aR$, with $l^0$ (chosen individually per input wire) and $R$ (circuit-wide constant) being vectors of random elements from $\mathbb{Z}_p$. We  call $l^0$ a \textit{base label}. The construction considers wires with different moduli, called \textit{mixed-moduli circuits}, and leverages different \textit{offset labels} $R_p$ for each module $p$. 
This allows for a Free-XOR generalization and free addition-gates as shown, e.g., in \cite{malkin2015whole}. Point-and-permute is generalized by using an element from $\mathbb{Z}_p$ instead of a single bit and choosing $1 \in \mathbb{Z}_p$ as the point-and-permute component of each base label $R_p$. A mixed-modulus-circuit consists of an acyclic structure of wires together with their moduli and gates constructed as follows:
\begin{itemize}[noitemsep,topsep=0pt]
  \item \textbf{Unbounded Fan-In Addition} To compute $(a,b)\mapsto (a + b) \bmod p$, we set 
  $l_x^a + l_y^b \equiv (l_x^0 + l_y^0) + (a + b)R_p \bmod p$.
  \item \textbf{Multiplication by a public constant} To compute $a \mapsto ac \bmod p$, we compute 
  $c \cdot l^a \equiv c \cdot l^0 + caR_p \bmod p$ with $c$ being co-prime to modulus $p$ (needed for technical reasons in the security proof, see also \cite{DBLP:conf/ccs/BallMR16}).
  \item \textbf{Projections for unary functions}  To compute an arbitrary unary function $\varphi\colon \mathbb{Z}_p \to \mathbb{Z}_q$, we construct
  a garbled projection gate with value $a$ on input wire $x$ that consists of $p$ ciphertexts of the form $\text{En}_{l_x^a}(l^{\varphi(a)})$. Note that leveraging a generalization of the garbled row reduction \cite{DBLP:conf/sigecom/NaorPS99}, one ciphertext could be removed. For simplicity, we ignore this optimization, as the saving is negligible for larger moduli.
\end{itemize}

While the first two gate types are free, the projection gate is not practical for large moduli $p$. Therefore, Ball \etal~use a \textit{composite primal modulus} (CPM) $P_k = 2 \cdot 3 \cdot \ldots \cdot p_k$ which is the product of the first $k$ primes. They leverage the Chinese remainder theorem to represent the label values in a \textit{residue (or CRT) representation}. Using CRT \-re\-pre\-sen\-ta\-tions for optimizations in GCs was proposed before, e.g., in \cite{DBLP:conf/focs/ApplebaumIK11}. From a circuit perspective, every value of a conventional wire is now represented by a bundle of wires, with each of the wires in a bundle corresponding to one residue in the CRT representation and one prime factor of the CPM $P_k$. 
Hence, instead of having a projection gate with $P_k$ possible inputs, we can now represent this projection gate via $k$ gates where the $i$-th gate has only $p_i$ possible inputs. Both addition and multiplication (by a constant) gates remain free.

\paragraph{Sign Gadget}
Leveraging the mixed-modulus HG (see \autoref{sec:mixed_mod_half_gate}) with new optimizations targeting ANN operations, Ball \etal \cite{DBLP:journals/iacr/BallCMRS19} demonstrate the practicability of garbled ANN inference. They propose a new approximated garbled $\operatorname{sign}$ gadget  (see also \autoref{sec:approximated_garbled_sign}) working over CRT representations with $\operatorname{sgn}_{l,h}(x)=l$ for $x \leq 0$ and $\operatorname{sgn}_{l,h}(x)=h$ for $x > 0$.

We use the sign function to construct the base-extension of our new scaling gadget. Following Ball \etal \cite{DBLP:journals/iacr/BallCMRS19}, \schemename{} uses the sign gadget to garble the $\operatorname{sign-}$ and $\operatorname{ReLU}$ activations and provide highly parallel GPU implementations. To garble the $\operatorname{sign}$ activation, the gadget is applied on the inputs $\operatorname{sgn-act}(a) = \operatorname{sgn}_{-1,1}(a)$. The $\operatorname{ReLU}$ activation is garbled via $\operatorname{ReLU}(a) = a \cdot \operatorname{sgn}_{0,1}(a)$, using the mixed modulus half gate for the multiplication.

\subsection{Trusted Execution Environments}
\label{subsec:trustedexecutionenvironments}
TEEs such as Intel SGX~\cite{DBLP:conf/isca/McKeenABRSSS13, DBLP:journals/iacr/CostanD16, sgx_dev_ref}, AMD SEV~\cite{kaplan2016amd}, Sanctum~\cite{DBLP:conf/uss/CostanLD16}, ARM CCA~\cite{DBLP:conf/osdi/LiLDGNSS22} and Intel TDX~\cite{DBLP:journals/access/SardarMF21} allow programs, containers, or whole VMs to be executed securely in hardware-sealed \emph{enclaves}. 
The hardware isolates the enclave from other programs on the host, regardless of privilege level or CPU mode. 
Enclave code and data are protected even from a compromised operating system or hypervisor.

The key feature of TEEs besides strong memory isolation is \emph{remote attestation}, a process that guarantees that a specific enclave, specific in the sense of the code and data it contains, has been deployed on a protected system. One distinguishes between local and remote attestation.
Using remote attestation, an enclave authenticates itself (i.e., the code, identity, and the fact it is executed in a trusted environment) to a remote third party. 
This makes TEEs particularly attractive for use in cloud computing offerings like MLaaS. 
The cloud provider creates an enclave to which the cloud customer establishes a secure connection and authenticates the enclave via remote attestation. 
If two or more distrustful parties want to perform outsourced computation together, they can outsource their data and computation to a mutually trusted enclave~\cite{DBLP:journals/scn/ChoiB19}.

Like \slalom{}, we use Intel SGX over newer TEEs such as TDX or AMD SEV to minimize the trusted computing base. 
SGX applications are divided into a trusted and untrusted component. 
The trusted component runs inside the enclave and communicates with the untrusted part outside the enclave through an interface defined by the developer. 
The trusted part of the application and the communication between the two components should be kept small to reduce the potential attack surface and improve performance.

Intel SGX was introduced in Intel Core CPUs in 2015, supporting 128MB to 256MB of \emph{Processor Reserved Memory} (PRM) shared by all active enclaves. The original use in consumer devices was discontinued, but SGX remains a heavily supported feature in server-grade CPUs such as the 4rd Gen Intel Xeon Scalable server CPUs, providing enclaves with up to 1TB of protected memory. 
SGX's enclave memory is encrypted and integrity protected outside the CPU package. 
The memory management unit (MMU) decrypts and integrity checks data before loading it from memory into caches or registers.
Nearly all server-level machines support TEEs (SGX by Intel or SEV by AMD), and they are used in practice, e.g. by the Signal messenger servers. As the main motivation for \schemename{} are MLaaS scenarios in which the sensitive computation is performed on a server level device, it is valid to assume a TEE can be used.

\section{Design of \schemename{}}
\schemename{} provides an efficient implementation of GCs for BMR circuits that outperforms previous approaches significantly and allows larger models to be garbled effectively. Improvements are achieved through three main contributions.  
Firstly, the implementation contains state-of-the-art techniques for GCs (see \autoref{sec:prelim}) and makes use of the approximated $\operatorname{ReLU}$ technique of Ball \etal \cite{DBLP:journals/iacr/BallCMRS19}, which gives a significant speedup.
 
Secondly, we make use of specialized ML hardware such as GPUs.
Previous implementations of GCs for ANNs, such as those by Ball et~al., use a \emph{scheduler} to choose the next gate in the GC to be evaluated. 
Due to this strictly linear approach, the massive parallelity of GPUs could not be used properly.
In contrast, we do not look at individual gates but treat gates on the same  layer as \emph{tensors}, i.e., as multidimensional objects called \emph{LabelTensors}.
This allows the evaluator to evaluate a complete circuit layer in parallel.
As a result, \schemename{} scales up easily to large circuits with very high width.
Classic examples of this kind of circuit are ANNs, but they also appear in many cryptographic use cases (e.g., via parallel encryption of blocks by \textsc{AES} or ChaCha20). 

Thirdly, \schemename{} includes a new garbled scale operation, allowing much larger networks to be processed due to being able to use more effective quantization schemes.

The modular approach taken in \schemename's implementation allows for many deployment scenarios.
For example, we can use TEEs both for the garbler and the evaluator, allowing the user to work with trusted data on an untrusted platform easily.
Our end-to-end framework supports the common interchange format ONNX (used with TensorFlow and PyTorch).
Hence, a user wanting to garble an ANN can easily enter a model in this format and will obtain a garbled version of it. 
This makes \schemename{} usable for everyone. 

\subsection{Quantization and Encoding}
ANNs typically operate over 32-bit floating point numbers. To increase the throughput and reduce the memory requirements of the models during the inference phase, models and inputs are quantized to integers. 
To garble ANNs, we need to quantize them as well. 
\schemename{} supports two quantization schemes we call \emph{SimpleQuant} and \emph{ScaleQuant}. 
SimpleQuant has been used before in GNNP \cite{DBLP:journals/iacr/BallCMRS19} and ScaleQuant is a more advanced quantization scheme, inspired by Gupta et~al. \cite{DBLP:conf/icml/GuptaAGN15} and also leveraged by \slalom{} \cite{DBLP:conf/iclr/TramerB19}. Using SimpleQuant to quantize a value $x$, we multiply with a small quantization constant $\alpha$ and round to the nearest integer: $\operatorname{SimpleQuant}(x, \alpha) = \operatorname{round}(x\cdot\alpha)$.

In ScaleQuant, we extend the forward-pass with down-scaling operations, which  allows quantizing larger ANNs without affecting the model's predictive power significantly. 
Without a garbled scale operation, previous arithmetic garbling schemes were not able to use ScaleQuant. This significantly limited the size of models that could be garbled effectively.
We introduce the garbled scale operation in \autoref{subsec:the-scaling-operation}.
To quantize a floating point number $x$, we use $\operatorname{ScaleQuant}(x, \ell) = round(x\cdot 2^\ell)$. In an ANN all weights are quantized with $\operatorname{ScaleQuant}(x, \ell)$ and all biases with $\operatorname{ScaleQuant}(x, 2\ell)$. In the forward pass, the outputs of layers with quantized parameters are then down-scaled with $\operatorname{ScaleQuant}(x, -\ell)$.

Our arithmetic GCs operate on values from the finite ring $\mathbb{Z}_{P_k}$. To be able to garble the quantized ANNs, all integers, including all weights, biases, and inputs, must be encoded in $\mathbb{Z}_{P_k}$. We map all positive integers, including zero, to the \emph{lower} half of $\mathbb{Z}_{P_k}$ and all negative numbers to the \emph{upper} half of the ring: $\operatorname{Encode}(x) = x \bmod P_k$ (see also \autoref{fig:scaling}).

\subsection{The Scaling Operation}
\label{subsec:the-scaling-operation}
Scaling a number $x$ by some \emph{scaling factor} $s$ means to compute $y= \lfloor x/s\rfloor$.
We only consider $s=2$, as successive applications of scaling by $2$ can be used to implement ScaleQuant.
First, we describe our scaling operation from a  high-level perspective without considering that the inputs to \schemename{} are in CRT representation.
Next, we explain the details regarding the CRT representation and how the scaling operation is implemented in \schemename.

\subsubsection{High-Level Steps}
\begin{figure}
    \centering
    \begin{subfigure}[t]{0.5\textwidth}
        \centering
        \includegraphics[width=0.95\linewidth]{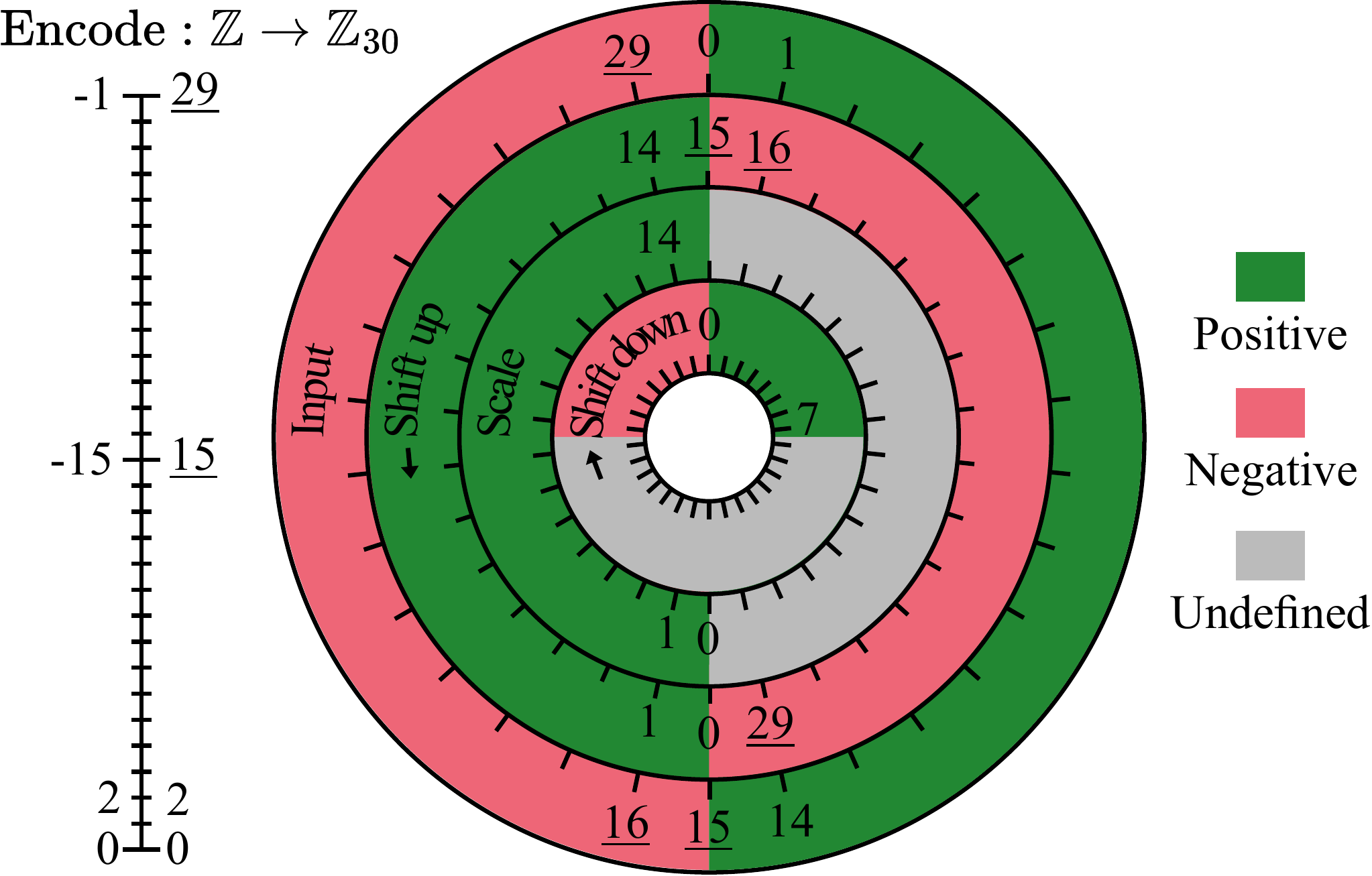}
        \caption{Schematic visualization.}
        \label{fig:scaling}
    \end{subfigure}%
    \hfill
    \begin{subfigure}[t]{0.5\textwidth}
        \centering
        \includegraphics[width=0.95\linewidth]{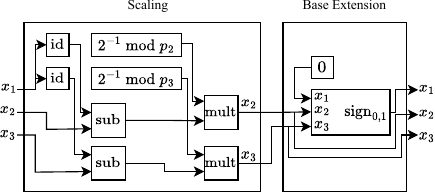}
        \caption{Circuit of the scaling gadget in \schemename.}
        \label{fig:scaling_crt}
    \end{subfigure}
    \caption{Visualization of our scaling operation in $\mathbb{Z}_{P_3}$. 
    }
\end{figure}

\autoref{fig:scaling} shows the steps (1) \emph{ShiftUp}, (2) \emph{Scale}, and (3) \emph{ShiftDown} of our scaling operation exemplarily for the case $P_3$. \autoref{tab:scaling_example} shows an example for $P_2=6$ including the outputs of all intermediate steps. The $\operatorname{ShiftUp}(x) = x + (P_k/2)$ step limits the result of the following $\operatorname{Scale}(X)=\lfloor x/2\rfloor$ step (including the base extension) to the positive value range $[0,P_k/2-1]$. Finally, we apply the $\operatorname{ShiftDown}(x) = x - \lfloor P_k/4\rfloor$ step to move the result to the correct value ranges $[0,\lfloor P_k/4\rfloor]$ and $[\lceil P_k/(3/4)\rceil,P_k-1]$, respective $[-\lfloor P_k/4\rfloor,\lfloor P_k/4\rfloor]$.

\begin{table}
    \centering
    \caption{Step-by-step outputs of the scaling gadget for all inputs in $\mathbb{Z}_{P_2}$. $x$: Input value to the scaling function. $x^\pm$: Sign information of $x$. $x^\uparrow$ and $x^\downarrow$: Output of the $\operatorname{ShiftUp}$ and $\operatorname{ShiftDown}$ operations. $b$: Scaling operation after the $\operatorname{ShiftUp}$ and before the base extension. $\varphi(x)$: Residue representation of $x$. $\varphi^{-1}()$ leverages the CRT to reconstruct a value from its residue representation. $y$: Output of our base extension algorithm.}
    \label{tab:scaling_example}
    \setlength\tabcolsep{2.9pt}
    \begin{tabular}{cccccccccccccc}
    $x$ & $x^{\pm}$ & $\varphi(x)$ & $x^\uparrow$ & $x^{\pm \uparrow}$ & $\varphi(x^{\uparrow})$ & $\alpha = [0, b(\varphi(x))]$ & $\varphi^{-1}(\alpha)$ & $\varphi^{-1}(\alpha)^\pm$ & $y$ & $\varphi(y)$ & $\varphi(y^\downarrow)$ & $y^\downarrow$ & $y^{\pm\downarrow}$ \\ \hline
    0 & 0 & [0, 0] & 3 & -3 & [1, 0] & [0, 1] & 4 & -2 & 1 & [1, 1] & [0, 0] & 0 & 0 \\
    1 & 1 & [1, 1] & 4 & -2 & [0, 1] & [0, 2] & 2 & 2 & 2 & [0, 2] & [1, 1] & 1 & 1 \\
    2 & 2 & [0, 2] & 5 & -1 & [1, 2] & [0, 2] & 2 & 2 & 2 & [0, 2] & [1, 1] & 1 & 1 \\
    3 & -3 & [1, 0] & 0 & 0 & [0, 0] & [0, 0] & 0 & 0 & 0 & [0, 0] & [1, 2] & 5 & -1 \\
    4 & -2 & [0, 1] & 1 & 1 & [1, 1] & [0, 0] & 0 & 0 & 0 & [0, 0] & [1, 2] & 5 & -1 \\
    5 & -1 & [1, 2] & 2 & 2 & [0, 2] & [0, 1] & 4 & -2 & 1 & [1, 1] & [0, 0] & 0 & 0 \\
    \end{tabular}
\end{table}

\subsubsection{CRT Representation}
The shift steps generalize easily to the CRT representation, as addition and subtraction can be performed independently on the individual residues. \autoref{fig:scaling_crt} visualizes the scaling step (with base extension). To scale the residues $\tilde{x}_2, \ldots, \tilde{x}_k$ after the $\operatorname{ShiftUp}$ in step 2 we use a simple technique described by Jullien \cite{DBLP:journals/tc/Jullien78} and compute $\tilde{y}_i = (\tilde{x}_i - \tilde{x}_0)\cdot 2^{-1} \bmod p_{i}$, where $2^{-1}$ is the multiplicative inverse of $2\bmod p_{i}$ and $p_i$ is the $i$-th prime number.

To construct the base extension and compute $\tilde{x}_1$ we leverage the observation that $\tilde{x}_1$ either contributes exactly $0$ or exactly $P_k/2$ to the sum of the reconstruction of $\tilde{x}$, since $\tilde{x} \equiv \sum_{i=1}^k \alpha_i \cdot \tilde{x}_i \bmod P_k$ where $\alpha_1 = P_k/2$ and $\tilde{x}_1 \in \mathbb{Z}_2$. Since all values must be positive after the scaling operation (see ring 3 in \autoref{fig:scaling}), we can use our $\operatorname{sign}$ gadget $\operatorname{sign}_{0,1}$ to test whether $\tilde{x}_1=0$ or $\tilde{x}_1=1$ (see base extension in \autoref{fig:scaling_crt}).

\subsubsection{Garbling}
This scaling operation is surprisingly cheap in our setting. For $P_{k}/2$ and $P_{k}/4$ we introduce $k$ additional constant garbled inputs respectively (offline).
The additions and subtractions of the shift operations are free in terms of needed ciphertexts.
To compute $\tilde{y}_i$ for $i\geq 2$ in the scaling step, the number $\tilde{x} \bmod 2$ must first be projected into the space $\{0,\ldots,p_{i}-1\}$ via a projection gate with two ciphertexts (see also \autoref{fig:scaling_crt}).
Then, the computation $(\tilde{x}_i - \tilde{x}_0)$ can be performed by several free subtractions.
The multiplication with $2^{-1}\bmod p_i$ is a multiplication with a constant and thus also ciphertext-free.
To finish the garbling of the scaling operation we need to garble a single $\operatorname{sign}$ gadget for the base extension. If $m_1, \ldots m_t$ represent the mixed-radix base used in the approximation of the sign gadget and $p_1, \ldots, p_k$ is a base of $k$ primes used for the residue representation, the cost (in terms of number of ciphertexts) to garble the sign gadget is $\alpha = t \sum_{i=1}^{k}pi + 2k \sum_{i=2}^n m_i + 2n(k-1) + m_1$. Garbling the scaling gadgets results in $\alpha + (k-1)2$ ciphertexts.

\subsection{System-level Architecture}
\begin{figure}
	\centering
	\includegraphics[width=1\linewidth]{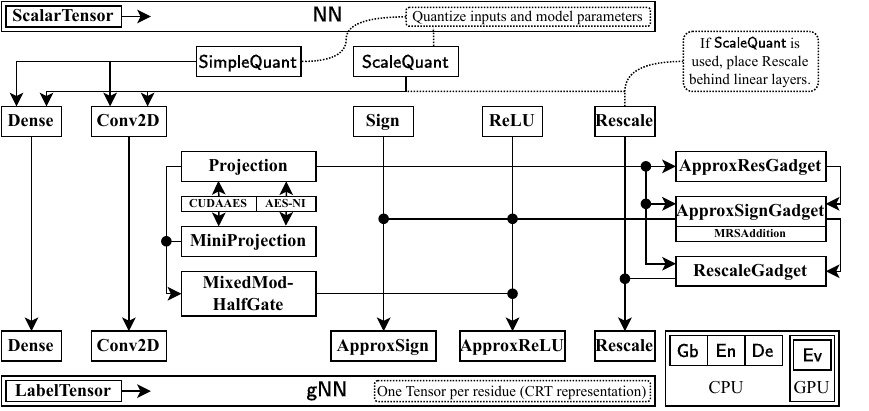}
	\caption{System-level architecture of \schemename{}.}
	\label{fig:system_level_architecture}
\end{figure}

\autoref{fig:system_level_architecture} shows a high level overview of \schemename's modular system-level architecture and components involved in the garbling process (from the top to the bottom). Garbling the Dense and Conv2D operations is straight-forward, as these linear operations only depend on multiplication with a public constant (the weight) and addition. To add the bias to the resulting label, we model it as a constant circuit input. To garble the ReLU activation, we follow the approach of Ball \etal \cite{DBLP:journals/iacr/BallCMRS19} see \autoref{sec:arithmetic_garbled_circuits} and \autoref{sec:appendix}. The garbling of the Rescale layer is performed as described in \autoref{subsec:the-scaling-operation}.

\subsection{Feature Set in Comparison}
\label{subsec:feature-comparison}
\newcommand*\rot[1]{\hbox to1em{\hss\rotatebox[origin=br]{-50}{#1}}}

\begin{table}[t]
  \centering
  \caption{Feature Comparison. CR: Communication rounds, CV: Communication volume.}
  \label{tab:featuresetcomparison}
  \setlength{\tabcolsep}{2.7pt}
  \scalebox{1}{
  \begin{threeparttable}
  \begin{tabular}{lcccccccccccccccccc}
    Scheme & \rot{\schemename{}} & \rot{\schemename{} (w/o TEE)} & \rot{\gnnp{} \cite{DBLP:journals/iacr/BallCMRS19}} &  \rot{\slalom{} \cite{DBLP:conf/iclr/TramerB19}} & \rot{\secureml{}  \cite{DBLP:conf/sp/MohasselZ17}} & \rot{\minionn{} \cite{DBLP:conf/ccs/LiuJLA17}} & \rot{CryptoNets \cite{DBLP:journals/corr/abs-1811-09953}} & \rot{\delphi{} \cite{DBLP:conf/uss/MishraLSZP20}}  & \rot{\cryptgpu{} \cite{DBLP:conf/sp/TanKTW21}} & \rot{\cryptflow{} \cite{DBLP:conf/sp/0001RCGR020}} & \rot{\mptwoml{} \cite{DBLP:conf/IEEEares/BoemerCD0Y20}} & \rot{\piranha{} \cite{DBLP:conf/uss/WatsonWP22}} & \rot{\simc{} \cite{DBLP:conf/uss/Chandran0OS22}} & \rot{\muse{} \cite{DBLP:conf/uss/LehmkuhlMSP21}} & \rot{\goten{} \cite{DBLP:conf/aaai/NgCWW021}} & \rot{\gazelle{} \cite{DBLP:conf/uss/JuvekarVC18}} & \rot{\deepsecure{} \cite{DBLP:conf/dac/RouhaniRK18}}\\\midrule
    TEE & \LEFTcircle & - & - & \CIRCLE & - & - & - & - & - &  \CIRCLE & - & - & - & - & \CIRCLE & - & -\\
    GPU &  \LEFTcircle & \LEFTcircle & - & \Circle & - & - & - & \Circle & \CIRCLE & - & - & \CIRCLE & - & - & \Circle & - & -\\
    Pure activation fun. & \LEFTcircle & \LEFTcircle & \LEFTcircle &  \CIRCLE & - & - & \LEFTcircle & - & \CIRCLE & \CIRCLE & \LEFTcircle & \CIRCLE & \CIRCLE & \CIRCLE & \CIRCLE & \CIRCLE & \Circle\\  
    Off-the-shelf models & \LEFTcircle & \LEFTcircle & \CIRCLE & \CIRCLE & \CIRCLE & \CIRCLE & \LEFTcircle & \LEFTcircle & - & \Circle& \CIRCLE & \CIRCLE & \Circle & \Circle & \CIRCLE & \CIRCLE & \Circle\\
    Convenient model-loading & \CIRCLE & \CIRCLE & \Circle & - & - & - & \Circle & \CIRCLE & \Circle & \CIRCLE & \CIRCLE & -  & - & - & - & - & -\\
    Constant CRs & \CIRCLE & \CIRCLE & \Circle & - & - & - & \CIRCLE & - & - & - & - & - & - & - & - & - & -\\
    Constant CV & \CIRCLE & \CIRCLE & \Circle & - & - & - & \CIRCLE & - & - & - & - & - & - & - & - & - & -\\
    Malicious security &  \CIRCLE & \Circle & \Circle & \CIRCLE & - & - & \CIRCLE & - & - & \CIRCLE & - & - &\LEFTcircle &\LEFTcircle & \CIRCLE & - & -\\ 
    Supports learning  & - & - & - & - & \CIRCLE &  - & - & - & \CIRCLE & - & - & \CIRCLE & - & - & \CIRCLE & - & -\\ 
    Unlimited Input Owner(s) & \CIRCLE & - & - & \CIRCLE & - & \CIRCLE & \CIRCLE & - & - & - & - & \Circle & - & - & - & - & \CIRCLE\\
    \bottomrule
  \end{tabular}
  \begin{tablenotes}
    \item \hfil $\LEFTcircle \text{ Optional}$,  $\CIRCLE \text{ Full}$, $\Circle \text{ Limited}$, $- \text{ No support}$
  \end{tablenotes}
  \end{threeparttable}}
\end{table}

Here, we compare \schemename's features and properties to other DPI schemes for ANNs. \autoref{tab:featuresetcomparison} shows that \schemename{} compares positively to these other schemes concerning the provided feature set. We explain the compared features: An activation functions is pure if it is not modified to be suitable for MPC techniques. Off-the-shelf models are conventionally trained models without tuning for the DPI setting. Convenient model-loading means out-of-the-box model usage from frameworks like PyTorch or TensorFlow. CR abbreviates constant number of communications rounds and CV communication volume with respect to the number of layer-transitions — in the model architecture — from linear to non-linear and vice versa. The cryptographic building blocks from GNNP would also fully support Constant CRs and Constant CVs if the scheme supported the offline pre-processing model, and the implementation would sacrifice the streaming feature.

The categorization into Optional, Full, Limited and No support arises from the critical path through the inference process. If a scheme supports offline pre-computation, the categorization is based on the online phase. Optional support means that the Scheme can suppress a feature in favor of performance or other requirements. Thus, optional support is in general stronger than full support. Limited support means that a feature is only partially supported, e.g., \slalom{} and \goten{} can only accelerate the linear layers of a model on the GPU and must communicate with the TEE for all non-linear layers.

Many frameworks do not use TEEs, which may be since SGX (one of the most common TEEs) constituted a severe memory bottleneck until recently. This has led to paging artifacts and immense performance losses. However, in the recent versions of SGX, the enclave memory is much larger and paging becomes less of a problem. The security benefits of the TEE for \schemename{} are discussed in the following section. 

The GPU support is one of the outstanding features of \schemename{}: Both linear and non-linear layers can be accelerated on a GPU without breaking the underlying security guarantees. Furthermore, due to the introduction of \emph{LabelTensors}, \schemename{} strongly benefits from the massive parallelity of garbled CNNs. Activation functions can either be computed with full accuracy, or as in many other works, in an approximated fashion to speed up the computation. While many frameworks support off-the-shelf models, meaning models that were trained conventionally without the DPI setting in mind, \schemename{} goes a step further: Users can conveniently load their models in the ONNX format, which is widely supported by standard ML frameworks such as TensorFlow or PyTorch. Together with the GPU-support, \schemename{} can be seen as a drop-in replacement for conventional and insecure inference engines in many existing ML applications.

\schemename{} requires only a single round of online communication regardless of the depth of the ANN, or the number of alternating linear and non-linear layers. While \gnnp{} \cite{DBLP:journals/iacr/BallCMRS19} explicitly does not exploit this outstanding property in its implementation \emph{fancy-garbling} and streams the GC to the evaluator during the online phase, \schemename{} capitalizes on this property and thus significantly accelerates the online phase. \schemename{} can evaluate the garbled ANN layer by layer without being slowed down by network communication at each transition from linear to non-linear layer (or vice versa). For a fixed input size (sum of inputs from all input providers) and a fixed CPM, \schemename{} has a constant online communication volume regardless of ANN architecture or the number of input providers.

If TEE-Support like Intel SGX is provided and the use-case agrees with the TEE-assumption (see \autoref{sec:security_of_dash}), \schemename{} can optimize the OTs for communicating the garbled inputs to the garbling device away and can run the inference with more than two input providers. This distinguishes \schemename{} from secret-sharing-based approaches, which are inherently limited in this aspect as they have to communicate and store a share for each input provider. Furthermore, trough leveraging the TEE \schemename{} achieves malicious security and can thus enable the use of ML inference in security critical domains, where protocol participants are not trustworthy and semi-honest security is not sufficient.

If the hardware does not provide TEE support, then \schemename{} does not achieve security against malicious attackers. However, all other \schemename{} optimizations are preserved and \schemename{} remains secure in the semi-honest outsourced inference scenario. In this scenario, \schemename{} still beats the current SOTA scheme (see also \cite{DBLP:conf/sp/NgC23}) \gnnp{} for the outsourced inference case (\gnnp{} also does not consider network overhead due to OTs in the evaluation).

Like other schemes, \schemename{} does not support training, as this requires constant parameter changes, which does not fit well with offline pre-computation. Secure training with feasible computation requirements is an open research problem and not the focus of this work. 

\subsection{Use Cases}
\begin{figure}
	\centering
	\includegraphics[width=1\linewidth]{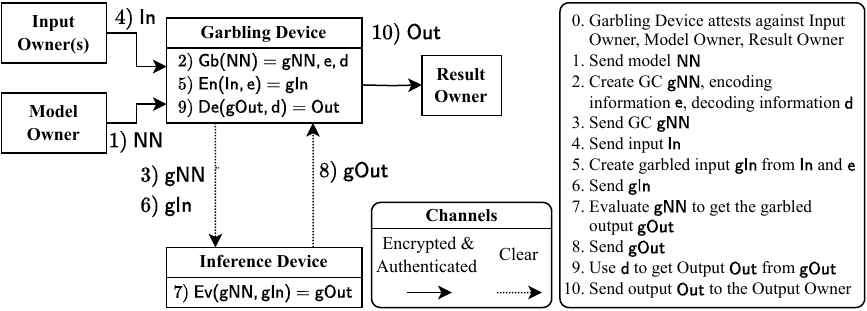}
	\caption{Example workflow of \schemename{}. The first four steps can be pre-computed in an input-independent offline phase. Note that the inference device works on garbled data.}
	\label{fig:workflow}
\end{figure}

Using \schemename{} guarantees several important security objectives, depending on the concrete scenario.
All scenarios share a common structure with the following participant roles:
\begin{itemize}[noitemsep,topsep=0pt]
    \item The \emph{model owner} who holds the ANN $\mathsf{NN}$. 
    \item The \emph{input owner} who holds an input $\mathsf{In}$ to the ANN (for inference). Our framework supports multiple input owners that each contribute a part of the input.
    \item The \emph{inference device}, a device aimed to compute the inference of an ANN.
    \item The \emph{garbling device}, a device to garble the circuit and the inputs.
    \item The \emph{result owner} who receives the inference result.
\end{itemize}
Dash runs in the typical outsourced inference scenario:  A customer outsources its inference workload to a single server in a fire-and-forget manner: Sending input and output is the only communication during the computation in the online phase. Dash allows fine adjustments to the participant roles in this setting. 
Some examples of conceivable scenarios are:
\begin{itemize}[noitemsep,topsep=0pt]
\item The classical outsourced inference scenario, where the model owner also controls the garbling device, the input owner also obtains the result as result owner, and the inference device is its own party. 
\item A more involved outsourced inference scenario with two different input owners.
\item A disjoint setting where each participant is its own entity. 
\end{itemize}
Compared to previous works like \slalom{}, \goten{} and the implementation \emph{fancy-garbling} of \gnnp{}, we do not assume a co-location of different devices or parties (especially the garbling and the evaluation device must \emph{not} be co-located). The typical workflow of \schemename{} is illustrated in \autoref{fig:workflow}.

Similar to Slalom~\cite{DBLP:conf/iclr/TramerB19} and \gazelle{}~\cite{DBLP:conf/uss/JuvekarVC18}, we also split the computation into an \emph{offline} and an \emph{online phase}.
In the offline phase (steps 1, 2, and 3), the model owner can already use the garbling device to prepare $\mathsf{gNN}$. Since this is typically the bottleneck of the computation, moving it to the offline phase allows us to speed up the online phase significantly. 
In the online phase, the input owner performs the inference with its sensitive data $\mathsf{In}$ and profits of the pre-computation. 
We assume the messages in step 1, 4, and 10 to be sent over an authenticated and encrypted channel. The messages in step 3, 6 and 8 are protected through the garbling properties.

\subsection{Security Objective}
We distinguish three security objectives: \emph{input privacy}, \emph{integrity of the inference}, and \emph{output privacy}. 
For many applications, the most sensitive information is the input $\mathsf{In}$. 
Following the notion of Tramèr and Boneh~\cite{DBLP:conf/iclr/TramerB19}, this security objective is called \emph{input privacy}. We always guarantee input privacy, even if all other participants cooperate (including other input owners).
Another essential property, called \emph{integrity of the inference}, ensures that, at the end of the protocol, the result owner really obtains $\mathsf{NN}(\mathsf{In})$, i.e., the inference is computed correctly. GCs inherently guarantee the integrity of the computation and the inputs~\cite{DBLP:conf/ccs/BallMR16}. 
However, in the case of a malicious model owner, we can't prevent the attacker from submitting a different network $\mathsf{NN}^{*}$. 
In this case, we will thus assume that the model owner first \emph{commits} to $\mathsf{NN}$ in order to allow verification that the correct network was used.
Analog to the input privacy, the output of the computation is also sensitive information. It should only be obtained by the result owner, so \emph{output privacy} can be guaranteed. We will always guarantee output privacy. 

While other works also aim to protect the model, many model extraction attacks (e.g., \cite{DBLP:conf/crypto/CarliniJM20, DBLP:conf/ccs/PapernotMGJCS17, DBLP:conf/uss/TramerZJRR16}) that reverse the architecture and the weights of a model from various sources, like input-output behavior, make model privacy an impossible goal for many scenarios. We thus do not provide it in our standard scheme. 
As we work in the outsourced inference setting, we always assume that the garbling device and the inference device do not collude.

\subsection{Security of \schemename{}}
\label{sec:security_of_dash}
Our security analysis against attackers is built on two assumptions: the \emph{garbling assumption}, which assumes that GCs are secure, i.e., they are private, oblivious, and guarantee authenticity~\cite{DBLP:conf/ccs/BallMR16}.
More formally, privacy guarantees that an 
attacker can not learn anything from $(\mathsf{gNN}, \mathsf{gIn},d)$ except for the correct output $\mathsf{Out}$.
Obliviousness means that $(\mathsf{gNN}, \mathsf{gIn})$ do not reveal information about $\mathsf{In}$.
Finally, authenticity means that an attacker knowing $(\mathsf{gNN},\mathsf{gIn})$ can not generate $\tilde{\mathsf{gOut}}\neq \mathsf{gOut}$ such that $\mathsf{De}(\tilde{\mathsf{gOut}},\mathsf{d})\neq \bot$.
A long line of research has established that one can base this assumption on different well-established cryptographic assumptions.
While this assumption guarantees security against a wide range of attacks, it does not prevent all attacks performed by malicious parties.
To prevent the remaining attacks, we use our second assumption, called the \emph{device assumption}, which guarantees that the garbling device acts as a trusted third party. 
After our security analysis, we will present two approaches on how to obtain the device assumption: one will be based on a TEE and the other will use purely cryptographic means.
We do not make additional assumptions about the system or the behavior of other parties.

A quick inspection of Figure~\ref{fig:workflow} shows that the security guarantees of GCs along with the device assumption guarantee security: 
Regarding the input (and output) privacy, by the device assumption, we only need to consider a scenario where some input owners, the model owner, and the inference device collaborate maliciously to obtain information about the input of an input owner or the output of the computation, as the garbling device acts as a trusted third party. The confidentiality of each input follows directly from the privacy property of GCs and 
the fact that the garbling device is trusted. Hence, given $\mathsf{gNN}$ and garbled inputs $\mathsf{gIn}$, an attacker is unable to learn anything about the plain input.
Similarly, nothing can be learned about the plain output without the decoding information $\mathsf{d}$ due to the obliviousness of the garbled circuit. Due to the device assumption, it is only sent to the result owner.

If the correct ANN is handed to the trusted garbling device, computation integrity again follows from the authenticity property of GCs and the fact the garbling device is trusted. An attacker cannot generate a manipulated output whose decoding is a valid output given the GC and garbled inputs.
Hence, the computation result will always be an encryption of $\mathsf{NN}(\mathsf{In})$, as the garbling device is only able to work on the garbled inputs $\mathsf{gIn}$ presented by the input owner and the $\mathsf{gNN}$.
We only need to guarantee that the model owner does indeed hand out the correct model to the garbling device. 
As discussed above, we ensure that the model owner first needs to commit to $\mathsf{NN}$, e.g., by publishing a hash of it.
As the garbling device is trusted due to the device assumption, after having obtained the model, it can verify that the given network matches the commitment.

\subsubsection*{On the Device Assumption}
In our security discussion above, we used the device assumption that guaranteed that the garbling device behaves honestly. We now discuss two ways to guarantee such a behavior.

In the first, purely cryptographic, way we consider the classical setting where the garbling device and the input owner are one entity. 
Input privacy and output privacy are trivially maintained by the privacy and the obliviousness of the GC. 
Finally, in order to prevent the garbling device from manipulating the circuit, we can either use the cut-and-choose approach~\cite{DBLP:conf/crypto/Chaum82,DBLP:journals/joc/LindellP15}, make use of zero-knowledge proofs~\cite{DBLP:conf/focs/GoldreichMW86}, or use authenticated garbling schemes~\cite{DBLP:conf/crypto/NielsenNOB12,DBLP:conf/ccs/WangRK17}. 
For example, when using zero-knowledge proofs, all parties first need to commit to their private information (such as the inputs or the model) and to their randomness. 
Whenever a party now send some value to another party, it also sends a zero-knowledge proof along to convince the other parties that the sent message is consistent with the protocol, the earlier messages, the committed private information, and the committed randomness.   
While such approaches have been dismissed as purely theoretical in the past, the last decade has seen the design of very efficient and small zero-knowledge proofs, see e.g., ~\cite{DBLP:conf/sp/BunzBBPWM18,DBLP:journals/dcc/AmesHIV23,DBLP:conf/ccs/YangSWW21}. 
These developments have been used to show that this approach does in fact allow for a relatively efficient implementation~\cite{DBLP:conf/ccs/AbascalSHIV20}.
To guarantee a private exchange of the input and output information, \emph{oblivious transfer} protocols~\cite{DBLP:journals/iacr/Rabin05} need to be used. 
For a more in-depth discussion see~\cite{DBLP:journals/ftsec/EvansKR18}.

The second approach uses a TEE on the garbling device and assumes that it is secure.
This assumption has been used already in different works, e.g., by Tramèr and Boneh~\cite{DBLP:conf/iclr/TramerB19}.
While TEEs have been studied for a much shorter time period, there are certain robust designs that have withstood attacks, e.g.~\cite{DBLP:conf/uss/CostanLD16, DBLP:conf/ndss/BrasserGJSS19}.
We stress here that even for the systems already in production, like Intel SGX or AMD SEV, the vast majority of attacks does not concern the cryptographic black-box guarantees, but make use of side-channel attacks. While TEEs are conceptually secure, side-channel attacks affect the implementation. Similarly, cryptographic primitives are also conceptually secure, but can be affected from side-channel attacks, if the implementation is not hardened. In protocol development, it is generally assumed that the underlying cryptographic primitives are implemented securely. It is therefore a reasonable assumption that the TEE implementation used for Dash also does not suffer from side-channel leakage.
Due to the remote attestation feature, third parties can be assured that code and data inside an enclave are protected and that the code is executed as expected. 
However, this can be completely done in the offline phase and thus does not influence the performance of the online phase.

\section{Implementation}
\schemename{} is implemented in C/C++ and CUDA 12.2 and uses the TEE assumption to guarantee security against malicious attackers, i.e., we assume that the TEE on the garbling device acts as a trusted third party.
To parallelize CPU computations, we leverage OpenMP. 
We use the Linux Intel SGX SDK version 2.21 (with in-kernel drivers) to support the latest Intel Saphire Rapids Scalable Xeon CPUs for our TEE implementation. For high-quality randomness in the wire label generation, we use Intel’s Digital Random Number Generator (\textsc{DRNG}). As a driver for our random-permutation-engine, we use the hardware-accelerated AES-NI instructions on the CPU and the OpenSSL AES (ECB mode) implementation on the CUDA-enabled GPU with T-Tables in constant memory. To send the wire labels made of $\mathbb{Z}_p$ elements to the \textsc{AES}-based fixed-key permutation, they first have to be compressed to 128-bit chunks. Like Ball \etal \cite{DBLP:conf/ccs/BallMR16}, we use the Horner method for compression. The compression $\operatorname{compress}(l)$ of a label $l$ with $n$ components and wire modulus $q$ is simply computed as 
$\operatorname{compress}(l) = (\ldots (l_nq+l_{n-1})q+\ldots)q+l_1$.

\subsection{\schemename{}-as-a-Framework}

\begin{figure}
    \begin{subfigure}{0.49\textwidth}
        \centering
        \begin{lstlisting}[frame=single]
// 1. Step: Create circuit from model-file
auto circuit = load_onnx_model("path", q_method);
// Optional: Optimize quantization on data
circuit->optimize_quantization(crt_size, example_data);
// 2. Step: Garble quantized circuit, sign accuracy: `acc`
auto gc = new GarbledCircuit(circuit, crt_size, acc);
// Optional: Move GC to GPU
gc->cuda_move();
        \end{lstlisting}
        \caption{Garbling a CNN given as ONNX model-file.}
        \label{lst:garbling}
     \end{subfigure}\ \ \ 
    \begin{subfigure}{0.49\textwidth}
        \centering
        \begin{lstlisting}[frame=single]
// Step 1. Quantize input
auto q_in = quantize(input, q_method);
// Step 2. Garble input
auto g_in = gc->garble_inputs(q_in);
// Optional: Move q_in to GPU
auto g_dev_in = gc->cuda_move_inputs(g_in);
// Step 3: Evaluate GC on CPU
auto g_out = gc->cpu_evaluate(g_in);
// Or on GPU
gc->cuda_evaluate(g_dev_in);
auto g_out = gc->cuda_move_outputs();
// Step 4: Decode outputs
auto out = gc->decode_outputs(g_out);
        \end{lstlisting}
        \caption{Evaluation of a GC.}
        \label{lst:evaluation}
    \end{subfigure}
    \caption{C++-interface of \schemename{}}
    \label{lst:cpp_interface}
\end{figure}

From the perspective of a model owner, the only thing needed to use \schemename{} is an ANN in the standard ONNX format.
Then, the general procedure to garble the ANN is as follows, where encoding of weights and inputs into $\mathbb{Z}_{P_k}$ is handled automatically.
\begin{enumerate}[noitemsep,topsep=0pt]
	\item Import the model as a circuit to \schemename{} and quantize weights and biases using SimpleQuant or ScaleQuant. 
 If SimpleQuant is used, optionally optimize quantization of the circuit with representative example data (the quantization constant is chosen based on a given CRT base $P_k$ and the maximal computed value during the inference).
	\item Garble the quantized circuit (and optionally move it to a GPU).
\end{enumerate}

Independently of the chosen use case,  the following steps must be performed to infer the plain output from an inference with the GC.
\begin{enumerate}[noitemsep,topsep=0pt]
	\item Quantize the inputs to integers.
	\item Garble the inputs (optionally move them to a GPU).
	\item Evaluate the GC on the garbled inputs (optionally move the result back to the host).
	\item Decode the garbled outputs.
\end{enumerate}

Like ANNs, our circuits and GCs consist of layers such as $\operatorname{Dense}$, $\operatorname{Conv2d}$, $\operatorname{ReLU}$, $\operatorname{Sign}$, $\operatorname{Flatten}$, or $\operatorname{Rescaling}$. 
All operations are implemented in a garbled and non-garbled variant to facilitate testing, experimentation, and quantization tuning. 
In addition to the ONNX model loader, users of \schemename{} can also construct circuits directly in code with an intuitive interface inspired by the sequential models of PyTorch and TensorFlow.
The interface of \schemename's SGX implementation behaves analogously to the implementation without a TEE.

\subsection{LabelTensors}

\begin{figure}
    \centering
    \begin{subfigure}[t]{0.45\textwidth}
    \centering 
	\includegraphics[width=1\linewidth]{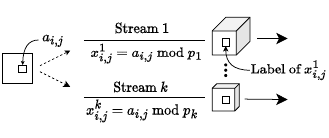}
	\caption{Leveraging LabelTensors in the CRT domain in independent streams on GPUs.}
	\label{fig:label_tensor}
    \end{subfigure}%
    \hfill
    \begin{subfigure}[t]{0.53\textwidth}
    \centering
	\includegraphics[width=1\linewidth]{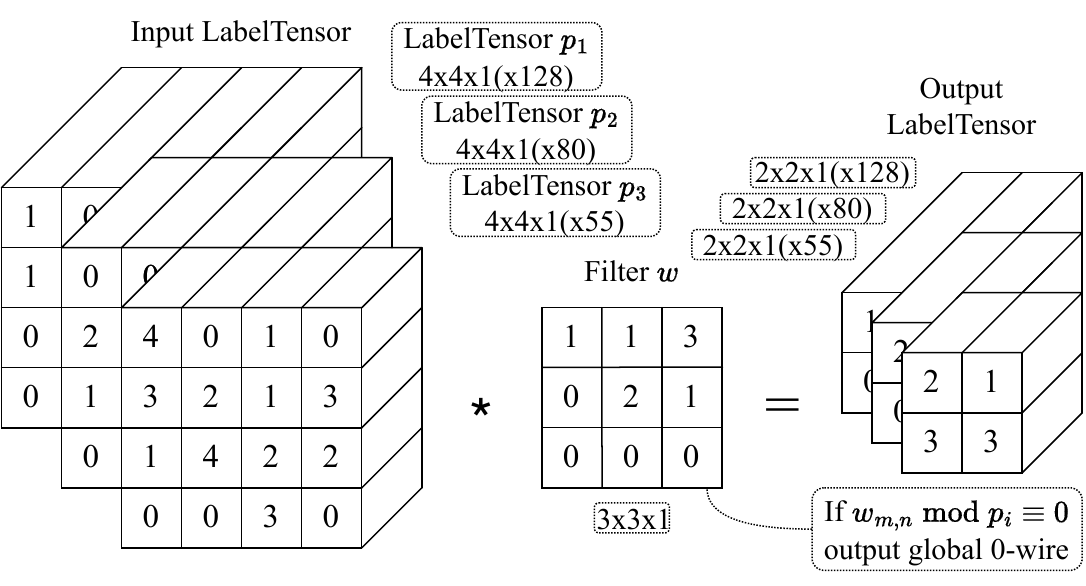}
	\caption{Conv2D operation over LabelTensors in Dash (for simplicity without bias, stride=1).}
	\label{fig:dash_conv2d}
    \end{subfigure}
    \caption{Visualization of the LabelTensor approach in Dash. The length $l$ of a label with modul $p_i$ is defined as $l=\left\lfloor 128 / \log_2(p_i)\right\rfloor$, since a longer label can not be packed into a 128bit chunk which is needed in the permutation used for garbling.}
\end{figure}

In general, arbitrary arithmetic GCs are not particularly well-suited for the evaluation on CUDA-enabled GPUs, as they do not fit nicely into the concept of single-instruction-multiple-thread (SIMT).
In CUDA threads are grouped into grids of blocks and blocks are executed in warps of 32 threads on the streaming multiprocessors of the GPU. As all threads of a warp share the same program counter, the parallel evaluation of heterogeneous gates, meaning the execution of threads with diverging control flows, will cause warp convergence, i.e. a serialization of those threads.
Furthermore, the arithmetic GCs handled by \schemename{} are neither strictly hierarchical nor uniform due to their global circuit wires and the modulus-based adaptive label length. To contribute enough entropy to the fixed-key AES permutation after packing, the labels with smaller modulus are longer (have more components) as the labels with larger modulus \cite{DBLP:conf/ccs/BallMR16}.
This makes memory coalescing difficult, since labels of gates that can be executed in parallel are not necessarily consecutively arranged in the global memory of the GPU.

Nevertheless, CNNs typically have quite large homogeneous layers consisting of many similar operations, which allows for optimizations with regard to a GPU evaluation. 
We introduce the notion of \emph{LabelTensors} to handle the labels in arithmetic GCs for CNNs efficiently. 
LabelTensors in \schemename{} are the counterpart of conventional tensors heavily used in machine learning frameworks like PyTorch or TensorFlow. 
The basic idea behind LabelTensors is to add another dimension to a conventional tensor to explicitly model the arithmetic GC labels.  
LabelTensors structure their values such that all labels of the same length lie along the tensor-width, -height and number of input-channels, consecutively in memory. Since all components of a label now lie consecutively in memory optimized CUDA kernels can coalesce memory accesses in warp granularity.

All garbled operations and utility functionalities  of our implementation, such as label en-/decryption and label de-/compression for CPU and GPU are based on these LabelTensors and work in the typical SIMT-style of CUDA without any further abstraction directly over the corresponding memory areas (see \autoref{fig:dash_conv2d} for a visualization of the Conv2D operation). For examples of the encountered operations over the LabelTensors see \autoref{subsubsec:model_architectures}.
This optimization concept removes the need for expensive gate-object scheduling.
In some sense, our approach can be seen as an arithmetical generalization of the JustGarble approach~\cite{DBLP:conf/sp/BellareHKR13}, which implements binary GCs with only memory blocks and indices and thus eliminates the need for expensive gate objects.
As opposed to the implementation of \gnnp{} (called \emph{fancy-garbling}) \cite{DBLP:journals/iacr/BallCMRS19}, which generates a large overhead due to gate-object scheduling with only eight threads, LabelTensors enable \schemename{} to efficiently leverage thousands of cores simultaneously on hardware accelerators like GPUs.
 
We use one LabelTensor per residue in the CRT representation and perform the operations on them in independent CUDA streams. This way, the operations over the residues of different CRT moduli and thus over labels of different length are processed in independent blocks on the streaming processors of the GPU and warp convergence is prevented (see also \autoref{fig:label_tensor}). The performance gain from the LabelTensor architecture was evaluated with several microbenchmarks, which will be presented in the next section. 

\section{Evaluation}
Intel offers a trusted library version of OpenMP \texttt{libsgx\_omp.a} for parallelization within SGX Enclaves. This library requires an OCALL and an ECALL to create a thread and an OCALL to wake up and pause threads and therefore has a considerable application-specific overhead compared to the conventional OpenMP library. For the model architectures in our evaluation, this overhead exceeds the performance gain achieved by our parallelization. Meaning, the usage of OpenMP inside the enclave is rendered useless for our model architectures.  Thus, \schemename{} only parallelizes CPU operations outside of the SGX enclave. However, the inference device is untrusted, and we can evaluate the inference in the online phase using the parallel CPU implementation without violating \schemename's security properties. The server features an Nvidia Geforce RTX 4090 GPU to evaluate \schemename's speedup.

\subsection{Microbenchmarks}
\begin{figure*}[t!]
    \centering
    \begin{subfigure}[b]{\linewidth}
        \centering
        \includegraphics[width=\linewidth]{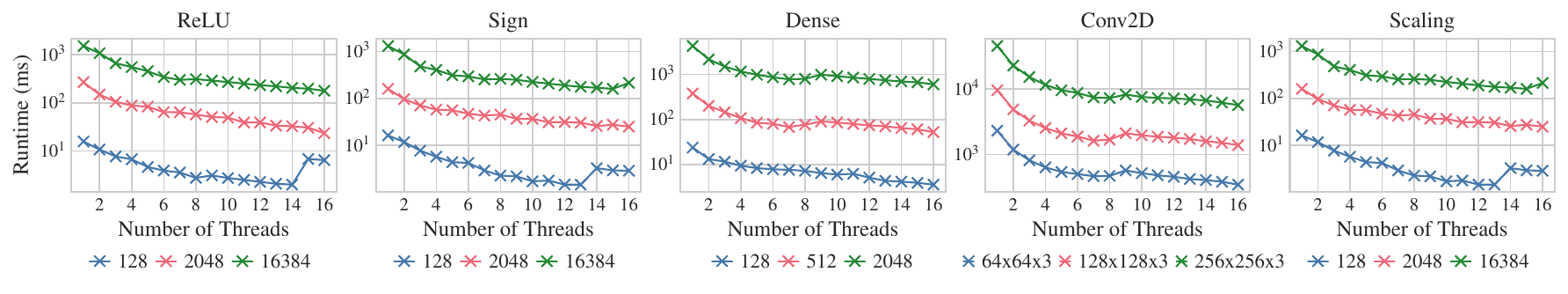}
        \caption{CPU-Scalability of \schemename. The $\operatorname{ReLU}$ and $\operatorname{Sign}$ functions are computed with 100\% accuracy.}
        \label{fig:microbenchmarks_scalability}
    \end{subfigure}
    
    \vspace{1em} % Adds space between the figures
    
    \begin{subfigure}[b]{\linewidth}
        \centering
        \includegraphics[width=\linewidth]{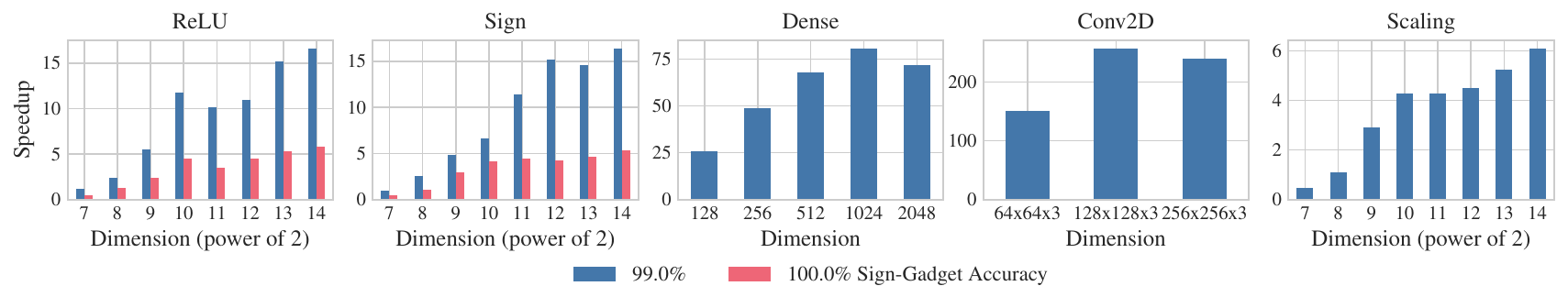}
        \caption{Speedup of \schemename's GPU against its CPU implementation.}
        \label{fig:microbenchmarks}
    \end{subfigure}
    
    \vspace{1em} % Adds space between the figures
    
    \begin{subfigure}[b]{\linewidth}
        \centering
        \includegraphics[width=\linewidth]{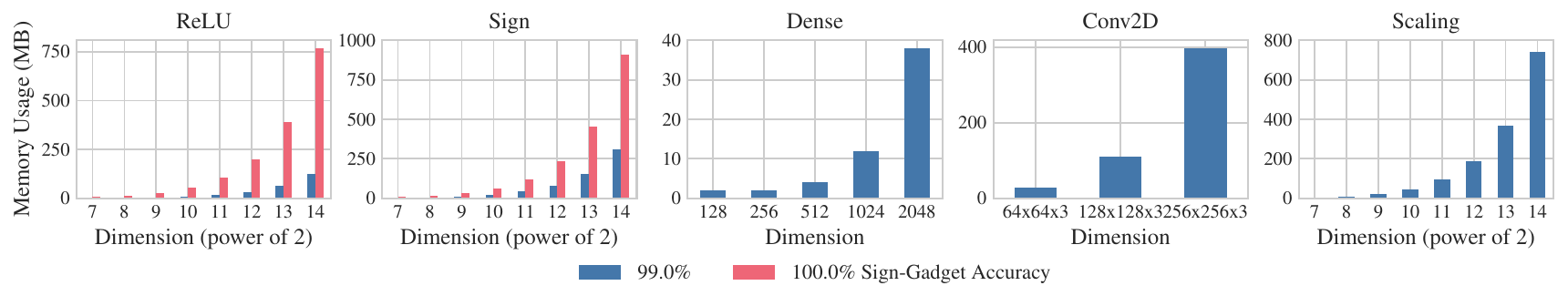}
        \caption{GPU memory usage after the GC and the garbled inputs are transferred to the device memory.}
        \label{fig:microbenchmarks_memory}
    \end{subfigure}
    
    \caption{Microbenchmarks of \schemename. Conv2D uses 16 filters of size 4x4 and a stride of 2.}
    \label{fig:combined_microbenchmarks}
\end{figure*}

For the microbenchmarks, we garbled all layers with CPM $P_8$, meaning the CRT representation consists of the residues modulo the first eight prime numbers. Furthermore, we excluded host-device transfer overheads since the hidden layers in CNNs do not suffer from them during the online phase. To evaluate the scalability of the garbled CNN layers over \schemename's LabelTensors on CPUs, we measured the runtimes for different input dimensions with 1-16 threads. The results in \autoref{fig:microbenchmarks_scalability} clearly show that every layer type, including the activation functions and our proposed scaling mechanism, benefit significantly from our parallelization-friendly LabelTensors. Apart from the transition from 8 to 9-11 threads (our CPU has eight physical cores), almost every additional thread brings a clearly measurable performance gain. The convolution operation benefits most from additional threads, but also, the other layers achieve speedups of up to an order of magnitude.

To evaluate the benefit of \schemename's GPU extension, we measure the achieved speedup against the CPU implementation, leveraging 16 threads. As shown in \autoref{fig:microbenchmarks}, all garbled layers, even the activation layers and the scaling layer, achieve a speedup when evaluated on the GPU. Starting with an input size of 256, our scaling layer can benefit from the GPU's parallelism and achieves a speedup of up to 6x for $2^{14}$ inputs. Above a dimension of 256 inputs, the GPU implementation of the activation functions shows a speedup-factor of 1.3 and 1.1 for the 100\% accurate ReLU respective sign activation functions. At $2^{14}$ inputs, the speedup grows to a factor of 5.8 respective 5.3, clearly outperforming the CPU implementation. Our implementation of the activation functions is based on the approximated $\operatorname{sign}$ gadget. For completeness, we also measured the speedup at a reduced precision. At 99\% precision, our GPU implementation can achieve a speedup of up to one order of magnitude. Since most of the computational effort in our garbled CNNs lies in the linear layers and we replaced max-pooling with strided convolutions compared to \gnnp{}, the speedup we achieve in CNN inference due to lower precision is negligible. In the following, our evaluation exclusively uses garbled activation functions with full precision. Nevertheless, the achieved speedup can make a difference if RAM-/GPU-Memory-Usage is a concern or in use cases in predictive modeling for logistic regression or small fully-connected ANNs. As expected, the garbled linear layers leverage the parallelism of the GPU best and achieve speedups of up to two orders of magnitude.

\autoref{fig:microbenchmarks_memory} shows the memory usage of \schemename's GPU implementation after the GC and the garbled inputs are transferred to the device memory. As expected, large inputs require the most memory, especially in the non-linear layers. If it is possible to compute with reduced accuracy, the memory requirement for the $\operatorname{ReLU}$ and $\operatorname{Sign}$ can be significantly reduced up to a factor of 7. To further reduce the compute and memory requirements, it would be interesting to explore the use of shorter CRT bases with larger moduli in future research. A shorter CRT base would result in less GC labels per circuit input and larger moduli in shorter labels with fewer components.

\subsection{Model Benchmarks}
\schemename{} supports the preprocessing paradigm, and we exclude the offline phase from our evaluation, as the cloud provider will pre-compute it before the customer data arrives.

\subsubsection{Model training} We trained all models with PyTorch over 100 epochs and selected the final model based on the minimal validation loss during the training. The models were trained on the MNIST and the CIFAR-10 data set. The MNIST data set consists of 70.000 black and white images of handwritten digits from zero to nine. 60.000 images serve as training data, and 10.000 images serve as test data. We excluded 5.000 images from the training data for validation. Each image is represented by 28 $\times$ 28 integer pixel values from the range [0,255]. The CIFAR-10 data set consists of 60.000 RGB-Images of size 32 $\times$ 32  with color values in the range [0, 255] and ten classes. There are 50.000 training images and 10.000 test images, and we excluded 5.000 images from the training data set for validation.

\subsubsection{Model architectures}
\label{subsubsec:model_architectures}
Similar to Ball et~al. \cite{DBLP:journals/iacr/BallCMRS19}, we evaluate several model architectures that have been used for evaluation of previous DPI frameworks \cite{DBLP:conf/sp/MohasselZ17, DBLP:conf/icml/Gilad-BachrachD16, DBLP:conf/dac/RouhaniRK18, DBLP:conf/ccs/LiuJLA17} (see \autoref{tab:model_architectures}). In some models, Ball et~al. replace $\operatorname{ReLU}$ activations with $\tanh$ activations during training and those $\tanh$ with $\operatorname{sign}$ activations at inference time to maintain model performance despite quantization. Thanks to our garbled scaling gadget and the ScaleQuant mechanism, we do not need such replacements. Springenberg and Dosovitskiy et~al. \cite{DBLP:journals/corr/SpringenbergDBR14} observed that replacing max-pooling with strided convolution in modern CNNs leads to competitive or even better predictive power. Following their approach enables us to heavily reduce the memory footprint of our garbled CNNs compared to \gnnp{} since garbling a single $\operatorname{max}(x,y)=x+\operatorname{ReLU}(y-x)$ operation results in the same large ciphertext overhead as garbling the $\operatorname{ReLU}$ function, while garbling the convolution is ciphertext-free.

MODEL-A to -D were trained on the MNIST and MODEL-f and -F on the CIFAR-10 dataset.
During garbling of the Models A-F we use CPMs $P_8, P_9, P_9, P_8, P_7, P_7$, such that there are no overflows in the GC during runs on the test sets. We used SimpleQunat for MODEL-A to MODEL-D and ScaleQuant with $l=5$ for MODEL-f and MODEL-F.

\begin{table}
    \setlength{\tabcolsep}{2pt}
	\centering
	\caption{Model architectures. $R$: ReLU, $(a)$: dense layer with $a$ outputs, $(a,b,c,d)$: 2d convolution with $a$ input-channel, $b$ output-channel, filter size $c$ and a stride of $d$.}
	\label{tab:model_architectures}
 \scalebox{1}{
	\begin{tabular}{ll}
	\toprule
	ModelA: & $(128), R, (128), R, (10)$\\
	ModelB: & $(1,5,5,1), R, (5,5,3,3), R, (5,10,3,1), R, (10,10,3,3), R, (100), R, (10)$\\
	ModelC: & $(1,5,4,2), R, (100), R, (10)$\\
	ModelD: & $(1,16,6,2), R, (16,16,6,2), R, (100), R, (10)$\\
	Modelf: & $(3,32,3,1), R, (32,32,3,1), R, (32,32,2,2), (32,64,3,1), R, (64,64,3,1), R,$\\
      & $(64,64,2,2), (64,128,3,1), R, (128,128,3,1), R, (10)$\\
	ModelF: & $(3,64,3,1), R, (64,64,3,1), R, (64,64,2,2), (64,64,3,1), R, (64,64,3,1), R,$\\
      & $(64,64,2,2), (64,64,3,1), R, (64,64,1,1), R, (64, 16, 1, 1), R, (10)$\\
	\bottomrule  
	\end{tabular}
}
\end{table}

\subsubsection{Model Performance}
\begin{figure*}
	\centering
	\includegraphics[width=\linewidth]{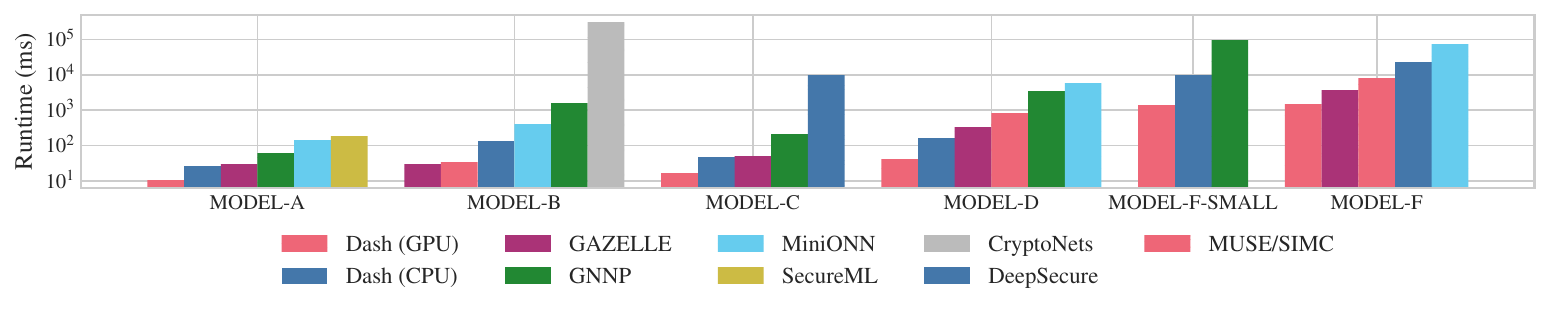}
	\caption{\schemename's online model runtime compared to previous DPI frameworks (data is not available for all framework/models, e.g. because not all frameworks support all models).}
	\label{fig:models_runtime_comparison}
\end{figure*}

\autoref{fig:models_runtime_comparison} compares the online runtime of \schemename{} to \gnnp{} \cite{DBLP:journals/iacr/BallCMRS19}, \minionn{}~\cite{DBLP:conf/ccs/LiuJLA17}, CryptoNets \cite{DBLP:journals/corr/abs-1811-09953}, \secureml{} \cite{DBLP:conf/sp/MohasselZ17}, \gazelle{} \cite{DBLP:conf/uss/JuvekarVC18}, \deepsecure{} \cite{DBLP:conf/dac/RouhaniRK18}, \muse{} \cite{DBLP:conf/uss/LehmkuhlMSP21} and \simc{} \cite{DBLP:conf/uss/Chandran0OS22} (\muse{} and \simc{} have the same online runtime). \schemename's runtimes incorporate the theoretically determined communication overhead in the worst case in a 1 GBit/s network under 100\% protocol overhead (see \autoref{sec:communication}). Achieving malicious security compared to semi-honest security is typically accompanied by a significant performance penalty. \schemename{} is, besides \gnnp{}, the only one of the given frameworks that achieves full malicious security and still delivers the best runtime performance.

Compared to \schemename, \secureml{} supports training. However, the evaluation of MODEL-A shows that \schemename{} is already over an order of magnitude faster in the case of small fully-connected ANNs. \deepsecure{} is also based on an optimized GC implementation, but in contrast to \schemename{} and \gnnp{}, it implements binary GCs. The results of MODEL-C show the advantage of arithmetic GCs for inference: \schemename{} is over two orders of magnitude faster than \deepsecure{}. While \minionn{} is significantly faster than CryptoNets, the evaluation of MODEL-B, -C, and -F clearly shows the speed advantage of \schemename.

\autoref{tab:model_runtime_comparison} shows a direct performance comparison of \schemename{} against \gnnp{} and \gazelle{}. \schemename's CPU and GPU implementation beat \gnnp{} concerning all model architectures. In the case of the small fully-connected MODEL-A, \schemename{} is 2x faster on the CPU and 6x faster on the GPU. For all other convolutional models, \schemename{} achieves a 14- to over 100-times speedup against \gnnp{}. Compared to \gazelle{}, \schemename's GPU implementations are always faster than \gazelle{}, except for MODEL-B.
For the largest evaluated MODEL-F, \schemename{} is over two times faster than \gazelle{}, showing its good scalability.

\autoref{tab:model_accuracy_comparison} reports the accuracy of our garbled models compared to \gnnp. \schemename{} beats \gnnp{} for each model. For the larger MODEL-f, the power of \schemename's ScaleQuant quantization becomes clear, \schemename's accuracy is almost 12\% higher compared to \gnnp{}.

\begin{center}
    \begin{minipage}[t]{0.49\textwidth}
        \centering
        \setlength{\tabcolsep}{2pt}
        \captionof{table}{Comparison of online runtimes (in ms). \gnnp{} does not incur communication overhead, \gazelle{} is not malicious security.}
        \begin{tabular}{lcccccc}
            \toprule
            Model & A & B & C & D & f & F \\
            \midrule
            CPU & 27 & 132 & 46 & 169 & 10263 & 23959 \\
            GPU & \textbf{10} & 32 & \textbf{16} & \textbf{42} & \textbf{1332} & \textbf{1443} \\
            \gnnp{} & 60 & 1520 & 210 & 3340 & 97000 & - \\
            \gazelle{} & 30 & \textbf{30} & 50 & 329 & - & 3560 \\
            \bottomrule
        \end{tabular}
        \label{tab:model_runtime_comparison}
    \end{minipage}%
    \hfill
    \begin{minipage}[t]{0.49\textwidth}
        \centering
        \setlength{\tabcolsep}{2pt}
        \captionof{table}{Comparison of the achieved model accuracy on the test-set in \% using \schemename{} and \gnnp{}.}
        \begin{tabular}{lccccc}
            \toprule
            Model & A & B & C & D & f \\ \midrule
            \schemename{} & \textbf{97.76} & \textbf{96.73} & \textbf{98.10} & \textbf{98.84} & \textbf{85.67} \\
            \gnnp{} & 96.80 & 86.72 & 97.21 & 96.44 & 73.74 \\ 
            \bottomrule
        \end{tabular}
        \label{tab:model_accuracy_comparison}
    \end{minipage}
\end{center}

\subsubsection{Runtime Distribution}

\begin{figure}
	\centering
	\includegraphics[width=1\linewidth]{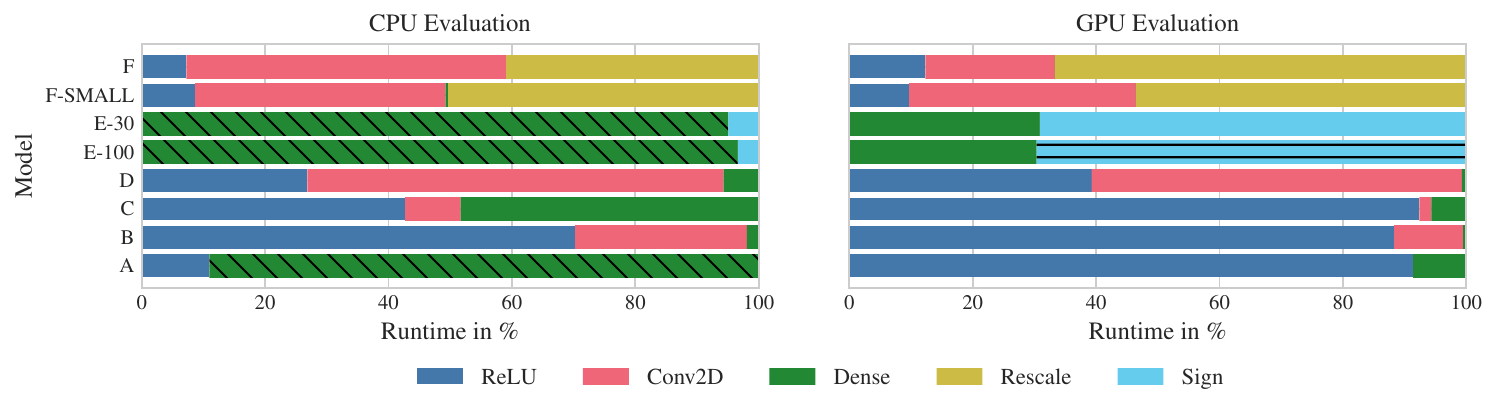}
	\caption{Runtime distribution over layer types and model architectures.}
	\label{fig:runtime_distribution_layer}
\end{figure}

\autoref{fig:runtime_distribution_layer} shows what portion of the runtime \schemename{} spends in which layer types. As expected, GPU acceleration shifts the distribution such that the linear portions are minimized, and the non-linear portions account for most of the runtime cost. For optimization, accelerating the activation functions on the GPU even further would be particularly interesting.

For larger models (see F models), the $\operatorname{ScaleQuant}$ mechanism is needed to maintain their predictive power. In this case, our garbled scaling dominates the runtime cost. Currently, our scaling gadget only supports scaling with two because otherwise, our base extension mechanism is not able to recover the lower residues. Since the F models use $\operatorname{ScaleQuant}$ with $l=5$ (rescaling by $2^{-5}$), the scaling gadget must be applied five times. For further research, it would be interesting to look for ways to enable base extension for residuals of larger CRT moduli to minimize the rescaling overhead.

\subsection{Communication}
\label{sec:communication}
We evaluate \schemename's online communication overhead in a theoretical model and assume the worst case, where the input owner(s), the garbling device, the inference device, and the result owner are separate parties that all communicate over a network. This way, we show that \schemename{} outperforms the given DPI schemes in all possible use cases. We incur costs for communicating the plain inputs $\mathsf{In}$, the garbled inputs $\mathsf{gIn}$, the garbled outputs $\mathsf{gOut}$, and the plain outputs $\mathsf{Out}$. We assume that all GC labels are compressed into 128-bit for communication. Plain inputs use a 16-bit, and plain outputs a 64-bit data type. For fairness, we assume a large protocol overhead of 100\% regarding the communication.

\begin{figure}
    \centering
    \begin{subfigure}[b]{0.45\textwidth}
        \centering
        \includegraphics[width=\linewidth]{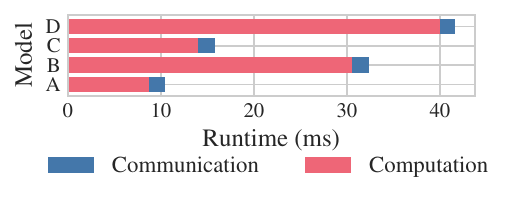}
        \caption{Distribution of the online workload.}
        \label{fig:distribution_communication_computation}
    \end{subfigure}%
    \begin{subtable}[b]{0.45\textwidth}
        \centering
        \caption{Online communication volume (MB) of \schemename{} and \gazelle{} (single input).}
        \setlength{\tabcolsep}{2pt}
        \begin{tabular}{lcccccc}
            \toprule
            Model & A & B & C & D & f & F \\ \midrule
            \schemename & \textbf{0.1} & \textbf{0.1} & \textbf{0.1} & \textbf{0.1} & \textbf{0.4} & \textbf{0.4} \\
            \minionn{} & 43.8 & 12.2 & - & 636.6 & - & 6226 \\
            \gazelle{} & 0.5 & 0.5 & 2.1 & 22.5 & 296 & - \\ \bottomrule
        \end{tabular}
        \label{tab:communication_cost}
    \end{subtable}
    \caption{Online Communication Volume.}
\end{figure}

\autoref{tab:communication_cost} compares \schemename's communication volume to \minionn{} and \gazelle{}. \schemename's online communication volume is constant for a constant CPM and input dimensions. For a CPM $P_k$, the volume of the garbled input is $k \cdot \text{size}(\mathsf{In}) \cdot 128$ bits, and for the garbled output $k \cdot \text{size}(\mathsf{Out}) \cdot 128$ bits. Typical DPI approaches that combine different MPC techniques for linear and non-linear layers require one communication round for each non-linear layer \cite{DBLP:conf/asiacrypt/ChangL01, DBLP:conf/mmsec/BarniOP06, DBLP:journals/ejisec/OrlandiPB07, DBLP:journals/tifs/BarniFLS011, DBLP:conf/sp/MohasselZ17, DBLP:conf/ccs/LiuJLA17, DBLP:conf/ccs/RiaziWTS0K18, DBLP:conf/ccs/MohasselR18, DBLP:conf/uss/MishraLSZP20}. Frequent communication slows down the computations because the inference device has to wait for a response before computing the next layer. \schemename{} requires only a single round of communication between the garbling device and the inference device, regardless of the depth of the ANN. These advantages can be seen in \autoref{fig:distribution_communication_computation}, which shows the distribution of communication and computation over the entire runtime. The share of communication work in the total runtime decreases with the model size from 15\% for MODEL-A over 5\% for MODEL-D to less than 1\% for MODEL-F.

\section{Conclusion}
We present a framework that offers ML inference with security against a malicious attacker by adopting optimized arithmetic GCs. The introduction of LabelTensors enables \schemename{} to efficiently accelerate the inherent parallelism of ANNs with parallel hardware such as multi-core CPUs and large, massively parallel GPUs.
Compared to \gnnp, leveraging a TEE allows \schemename{} to host inference applications with more than two input providers. With a fixed input dimension, the emerging communication volume and the memory requirement on the inference device remain
constant regardless of the number of input providers. Independent of the number of alternating linear and non-linear layers, \schemename{} requires only one round of communication between the garbling and the inference device. As a result, \schemename{} achieves state-of-the-art performance for all evaluated model architectures. \schemename{} can run existing models without retraining and does not require knowledge about GCs.

We exhibit a large feature comparison to many previous approaches and a thorough performance evaluation based on both micro-benchmarks and real-life models to show that \schemename{} outperforms previous works both in regard to resource consumption (runtime and communication) and feature set. This work provides a relevant step towards making secure inference fast and easy to use, which will aid a broader adoption of proposed protection mechanisms, resulting in making secure inference accessible to a wide public audience.

\section{Acknowledgments}
We thank the anonymous reviewers for their useful feedback. This work has been supported by the BMBF through the project AnoMed.

\section{Appendix}
\label{sec:appendix}
\paragraph{Half-Gate Generalization}
Ball \etal~\cite{DBLP:journals/iacr/BallCMRS19} introduced a mixed-modulus multiplication gate. We denote the labels on the input-wires $x,y$ with semantic $a,b$ by $l_x^a, l_y^b$ and the point-and-permute value of $l_y^0$ by $r$.
Hence, the point-and-permute value of $l_y^b$ is given by $(b + r) \bmod p$. The key idea is a reformulation of the multiplication into two gadgets $a \cdot (b+r)$ (the evaluator half gate) and $a\cdot r$ (the garbler half gate).
Since $r$ is the color value of the base label of the $y$-wire, the garbler already knows $r$ at garbling time, and the evaluator knows $b + r$ at evaluation time. 

The garbler selects a random base label $l_u^0$ for the garbler HG and performs $\operatorname{EN}_{l_x^0 + aR}(l_u^0 + arR)$ for all $a \in \mathbb{Z}_p$.
For the evaluator HG, the garbler selects a random base label $l_v^0$ and performs $\operatorname{EN}_{l_y^0 + bR}(l_v^0 - (b + r) l_x^0)$ for all $b \in \mathbb{Z}_p$.
The evaluator knows the labels of both inputs and decrypts two of the above ciphers accordingly. Knowing $b + r$ and $l_x^0 + aR$, he can choose $l_v^0 - (b + r) l_x^0 + (b + r) (l_x^0 + aR) = l_v^0 + (b + r)aR$ and the product-output is $l_v^0 + (b + r)aR - (l_u^0 + arR) = l_v^0 - l_u^0 + abR$ (considering $l_v^0 - l_u^0$ as the base label).

\paragraph{Mixed-Modulus Half-Gate}
\label{sec:mixed_mod_half_gate}

Ball \etal~\cite{DBLP:journals/iacr/BallCMRS19} show how to reduce the mixed modulus multiplication cost to roughly $q + p + 1$ ciphertexts. To encrypt the evaluator HG, the garbler uses an input label $l_y^0 + bR'$ with modulus $q$. Instead of $p$ ciphertexts, the evaluator HG results in $q$ ciphertexts. Since $b \in \mathbb{Z}_q$, the corresponding color value is also from $\mathbb{Z}_q$. Now, we have to preserve the known input $b + r \in \mathbb{Z}_p$ of the evaluator HG. Therefore, Ball \etal \cite{DBLP:journals/iacr/BallCMRS19} propose to leverage a single projection gate of $q$ very short ciphertexts, encrypting $b + r$ for all possible $b$. For the multiplication $\operatorname{ReLU}(a)=\operatorname{sgn}(a) \cdot a$, we can pack all of these ciphertexts into $128$ bits. Garbling the evaluator HG for all $b \in \mathbb{Z}_q$ results in the ciphertexts $\operatorname{EN}_{l_y^0 + b R'}(l_v^0 - (r + b) \cdot l_x^0) \text{ and } \operatorname{EN}_{l_y^0 + b R'}(r + b)$.

\paragraph{Approximated Garbled Sign}
\label{sec:approximated_garbled_sign}
The garbled $\operatorname{sign}$ function $\operatorname{sgn}: \mathbb{Z}_{P_k} \to \{0,1\}$ of Ball et~al. \cite{DBLP:journals/iacr/BallCMRS19} expects $\mathbb{Z}_{P_k}$-values and interprets the first half of the ring as negative and the other half as positive numbers, i.e., $\operatorname{sgn}(x)=0$ if $x < P_k/2$ and $\operatorname{sgn}(x) = 1$ if $x\geq P_k/2$.

The concept\footnote{This general approach also appears in earlier works like \cite{hung1994approximate}.} is based on the CRT, which describes the reconstruction of the value $x \in P_k$ from its residue representation $\llbracket x \rrbracket_{P_k} = (x_1, \ldots, x_k)$, where $x_i = x \bmod p_i$:
    $x \equiv \sum_{i=1}^k A_i^{-1} \cdot A_i \cdot x_i \bmod P_k \text{, with } A_i = \frac{P_k}{p_i}
      \equiv \sum_{i=1}^k \alpha_i \cdot x_i \bmod P_k$.
       Hence, the $\operatorname{sign}$ function can be computed just regarding the fractional part of the last summand by computing 
$\alpha_i x_i / P_k$ and rounding it to $1/M$ for some $M$. 
This approximation can be represented as $\mathbb{Z}_m$-wire and the the error of a $k$ term sum is limited by $k/2M$. Hence, for $M > kP_k/2$, the result is correct. In fact, this $\mathbb{Z}_m$-wire, this is represented as a bundle of wires using a mixed-radix representation (see \cite{DBLP:journals/iacr/BallCMRS19}).
In other situations, correctness is not needed, and $M$ is a trade-off parameter between precision and garbling cost.

\subparagraph{Garbled Mixed-Radix Addition}
For use in the approximated garbled $\operatorname{sign}$ function, Ball et~al. \cite{DBLP:journals/iacr/BallCMRS19} introduced a fast \textit{mixed-radix addition}. Consider the summation of $k=3$ values represented in mixed-radix representation $\mathbb{Z}_{D_n} \cong (\mathbb{Z}_{d_1} \times \ldots \times \mathbb{Z}_{d_n})$ (associated with the integers $\{0, \ldots, D_n - 1 =(\prod_i d_i)-1\}$ and $d_1$ being the most significant digit). The operation first computes $s = x + y + z + c_i^{\mathtt{in}}$, where $x,y,z$ are the values of the $\mathbb{Z}_{d_i}$-wires and $c_i^{\mathtt{in}}$ is the carry-input.
Then, $x,y,z,c_i^{\mathtt{in}}$ are cast using four projection gates to $\mathbb{Z}_{3d_i + c_i^{\mathtt{max}} - 1}$ and all input values are added $\bmod\; 3d_i + c_i^{\mathtt{max}}-1$.
Finally, the carry-out $c_{\mathtt{out}} = \lfloor \frac{x + y + z + c_i^{\mathtt{in}}}{d_i}\rfloor$ is computed using an unary gate.

\bibliographystyle{alpha}
\bibliography{biblio}

\end{document}